\documentclass[fleqn,usenatbib]{mnras}

\usepackage{newtxtext,newtxmath}

\usepackage[T1]{fontenc}
\usepackage{ae,aecompl}
\usepackage[usenames, dvipsnames]{color}

\usepackage{graphicx}	
\usepackage{amsmath}	
\usepackage{pdflscape}
\usepackage{multirow}
\usepackage{subfigure}

\newcommand{\cmps}{\ensuremath{\textrm{cm}~\textrm{s}^{-1}}}
\newcommand{\mps}{\ensuremath{\textrm{m}~\textrm{s}^{-1}}}

\newcommand{\mystar}{HD\;13808}

\newcommand{\lrhk}{\ensuremath{\log R'_{\rm HK}}}
\newcommand{\bis}{\ensuremath{{\rm BIS}}}
\renewcommand{\th}{\textsuperscript{th}} 

\newcommand{\ms}{\,m\,s$^{-1}$}
\newcommand{\cms}{\,cm\,s$^{-1}$}

\title[Two Neptune mass planets  orbiting HD\;13808]{The HARPS search for southern extra-solar planets XLV. \\ Two Neptune mass planets  orbiting HD\;13808: a study of stellar activity modelling's impact on planet detection\thanks{Based on observations made with HARPS spectrograph on the 3.6-m ESO telescope  at La Silla Observatory, Chile}}

\author[E. Ahrer et al.]{E. Ahrer$^{1,2}$\thanks{E-mail: eva-maria.ahrer@warwick.ac.uk},
D. Queloz$^{1,3}$,
V.~M.~Rajpaul$^{1}$,
D.~S\'egransan$^{3}$, 
F. Bouchy$^{3}$,
R. Hall$^{1}$,
\newauthor
W. Handley$^{1,4}$,
C. Lovis$^{3}$,
M. Mayor$^{3}$,
A. Mortier$^{1,4}$,
F. Pepe$^{3}$,
S. Thompson$^{1}$,
 \newauthor 
S. Udry$^{3}$, N. Unger$^{3}$
\\
$^{1}$Astrophysics Group, Cavendish Laboratory, JJ Thomson Avenue, CB3 0HE Cambridge, UK\\
$^{2}$Department of Physics, University of Warwick, Gibbet Hill Road, CV4 7AL Coventry, UK\\
$^{3}$Departement d'astronomie, Universit\'e de Gen\`eve, Chemin des Maillettes 51, CH-1290 Versoix, Switzerland\\
$^{4}$Kavli Institute for Cosmology, Cambridge, Madingley Road, CB3 0HA Cambridge, UK\\
}

\date{Accepted XXX. Received YYY; in original form ZZZ}

\pubyear{2020}

\begin{document}
\label{firstpage}
\pagerange{\pageref{firstpage}--\pageref{lastpage}}
\maketitle

\begin{abstract}
We present a comprehensive analysis of 10 years of HARPS radial velocities of the K2V dwarf star \mystar, which has previously been reported to host two unconfirmed planet candidates. We use the state-of-the-art nested sampling algorithm \textsc{PolyChord} to compare a wide variety of stellar activity models, including simple models exploiting linear correlations between RVs and stellar activity indicators, harmonic models for the activity signals, and a more sophisticated Gaussian process regression model. We show that the use of overly-simplistic stellar activity models that are not well-motivated physically can lead to spurious `detections' of planetary signals that are almost certainly not real. We also reveal some difficulties inherent in parameter and model inference in cases where multiple planetary signals may be present. Our study thus underlines the importance both of exploring a variety of competing models and of understanding the limitations and precision settings of one's sampling algorithm. We also show that at least in the case of \mystar, we always arrive at consistent conclusions about two particular signals present in the RV, regardless of the stellar activity model we adopt; these two signals correspond to the previously-reported though unconfirmed planet candidate signals. Given the robustness and precision with which we can characterize these two signals, we deem them secure planet detections. In particular, we find two planets orbiting \mystar\ at distances of $0.11, 0.26$~AU with periods of $14.2, 53.8$~d, and minimum masses of $11, 10$~$M_\oplus$. 

\end{abstract}




\section{Introduction} 
The Radial Velocity (RV) method has been an important and productive tool for discovering exoplanets ever since it led to the discovery of the first exoplanet orbiting a Sun-like star \citep{Pegasi51b}. However, the search for small Earth- and Neptune-like planets orbiting Sun-like stars is very challenging. They give rise to relatively small radial velocity signatures of order 1~\mps\ or less, whilst stellar magnetic activity can induce RV signals that can mimic planetary ones, with amplitudes of order many meters per second \citep[e.g][]{Didier2001}.

Thus, with the increasing precision of RV measurements facilitated by upcoming extreme-precision Doppler spectrographs such as  ESPRESSO \citep{ESPRESSO}, EXPRES \citep{EXPRES}, HARPS3 \citep{HARPS3}, HIRES \citep{CODEX2008} and NEID \citep{NEIDSchwab2016}, the characterization of the influence from host stars' activity on the RV measurements becomes ever more important. Hence there is a strong interest in developing tools to disentangle stellar and planetary signals in RV data.

A variety of methods exists that attempt to `correct' or model stellar activity signals in RVs, and thus reduce the possibility of false-positive planet detections. Models describing this include pre-whitening and red-noise models \citep[e.g][]{Hatzes2010,Feroz2013}, as well as studying stellar activity indicators such as the bisector inverse slope (BIS) and Full Width Half Maximum (FWHM) of the cross-correlation function (CCF) between a target star and a template spectrum, or measurements of chromospheric activity in the target star, such as the \lrhk\ index \citep[e.g][]{Boisse2009,Queloz2009, Dumusque2011}. The presence of periodic signals in indicators usually suggest an activity-induced signal rather than a planetary one, since a genuine planet would induce a periodic Doppler shift in all spectral lines, but would not produce the same periodic variations in the stellar activity indicators. It is straightforward to use linear correlations \citep[e.g.\ ][]{Didier2001} and to include harmonic models which use the rotational period of the star and its harmonics \citep[e.g][]{Dumusque2012} to describe stellar activity induced RV variations. 

Other, more computationally expensive methods include modelling the stellar surface features directly \citep[e.g][]{SOAP2012} and predicting activity-related RV variability with stellar activity indicators using the $FF'$ method and Gaussian processes (GPs) as described in e.g.\ \citet{Aigrain2012} and \citet{RajpaulGP}; such approaches have already successfully been used to identify false-positive planetary detections as e.g.\ in \citet{haywood2014} or \citet{rajpaul2016}.

To quantify the quality of different models for describing observed RV data, Bayesian model comparison has proven to be a robust approach \citep[e.g][]{Feroz2013,Faria2016,Richard2018}, also in combination with GP modelling \citep{Faria2020HD41248}. However, a significant challenge inherent in this method is very high computing costs when dealing with high-dimensional problems, since computing Bayesian model evidences entails integrating over all model parameter posteriors; a related challenge is degeneracy inherent in combinations of Keplerian and stellar activity models \citep[see][]{PlanetEvidence}. Nevertheless, it is widely accepted that Bayesian model comparison is a theoretically-sound approach to answering questions about competing physical models,\footnote{The same can \emph{not} be said for various other commonly-used approaches to model selection, such as residual minimization, Bayesian information criterion testing, etc.\ \citep{gelman2013bayesian}} and that such computational burdens are therefore a price worth paying \citep{goodman1999toward}. 

This paper presents a comprehensive analysis of a series of 246 spectroscopic measurements on \mystar\ carried out with HARPS from 2003 to 2014. This star belongs to a sample of bright stars selected for their low level of RV ``jitter'' to maximize the sensitivity of the survey to low-mass exoplanets. Series of stellar spectrum of \mystar\ with a signal to noise (SN) ratio large enough to reach the photon noise RV measurement error of order 50\cms\ are available. In addition, the exposure time was at least 15~minutes to minimize the effect of acoustic mode stellar oscillations \citep[e.g.\ ][]{dumusque2011granulationoscillations, Chaplin2019}. 
Early in the survey of HD 13808, small amplitude  RV variations were detected, suggesting a combination of time variable signal due to a multi-planet systems and moderate stellar activity. We present a comprehensive analysis of HARPS RVs on that star using Bayesian inference to combine various stellar activity models with a set of independent Keplerian orbit solutions.  

With this paper we aim to examine the importance of comparing systematically a number of competing physical models, since the usefulness of Bayesian model comparison is limited by the quality of the models considered (favouring one inappropriate model over another does not imply that either model is in any sense `correct'). We show that if an inadequate stellar activity model is considered, one may be led astray and end up with spurious conclusions about the number of planets present in one's system. Models we consider in our analysis include: linear correlations between RV data and stellar activity indicators; sinusoidal models; and joint GP regression of RV measurements and activity. Moreover, we present a thorough test of the nested sampling algorithm \textsc{PolyChord}, a state-of-the-art sampler, and discuss its limitations and the importance of the precision settings in exoplanet applications. The \mystar~system is used as a test case for this study as (i) the host star shows stellar activity variability, and (ii) it has been suggested in the literature that the system hosts at least two planet candidates, though the existence of these planets was never securely confirmed (see \citealt{Mayor2011} and \citealt{Gillon2017}; at the time of writing this paper, the NASA Exoplanet Archive\footnote{Available online at \url{exoplanetarchive.ipac.caltech.edu}.} lists \mystar\ as having no confirmed planets). With the analysis in this paper the status of these candidates elevates to confirmed planets. 

This paper is structured as follows. We describe the HARPS observations in Section~\ref{sec:harps_observations}, followed by the modelling of Keplerian orbits in Section~\ref{chapter:RV}. We introduce the stellar activity models applied in this work in Section~\ref{chapter:stellar} and outline model comparison using Bayesian inference and \textsc{PolyChord} in Section~\ref{chapter:Bayesian}. Afterwards, we present the results of our analysis of \mystar \ RVs in Section~\ref{chapter:results_HD13808} and conclude with our discussion in Section~\ref{chapter:discussion}.

\section{HARPS Observations of \mystar}
\label{sec:harps_observations}

Spectra of \mystar\ were measured with HARPS, a fiber-fed spectrograph installed in a vacuum vessel, mounted on the 3.6m ESO  telescope of La Silla  in Chile \citep{2003Msngr.114...20M}. Observations were made using the  most precise observing mode which utilizes a simultaneous calibration by the ThAr calibration lamp. In this configuration HARPS achieves a long term RV precision better than 1\ms\ allowing us to detect small stellar RV variations of this order \citep{2008ASPC..398..455L}.

\begin{table}
    \centering
     \caption{Description of the HARPS RV data of \mystar.}
    \label{tab:full_data_stats}
    \begin{tabular}{lcc}
    \hline
    \multicolumn{3}{c}{HD13808} \\ \hline
         Number of observations & \multicolumn{2}{c}{246} \\
         Time span (days) &  \multicolumn{2}{c}{4051} \\
         mean(RV) (\mps) & \multicolumn{2}{c}{41\,095 }  \\
         rms(RV) (\mps)& \multicolumn{2}{c}{3.94 } \\
         mean($\sigma_{\rm{RV}}$) (\mps)& \multicolumn{2}{c}{0.76} \\
         median(\lrhk) & \multicolumn{2}{c}{-4.90} \\ 
         \hline
    \end{tabular}

\end{table}

The data obtained by HARPS are automatically processed on site by a data reduction software -- the HARPS DRS --  that extracts the spectra, calibrates it and eventually computes a cross-correlation function with a stellar template. The stellar radial-velocity is measured from the CCF as well as  other parameters like the FWHM  or the BIS  \cite{Didier2001,2002A&A...392..215S}. In addition for each HARPS spectra,  the value of the Calcium S activity index is estimated from the chromospheric re-emission in the Ca II H and K lines and converted in standard \lrhk\ index.

Since the release  of the first version of the HARPS DRS in 2003, a series of successive versions with improved algorithms have been developed, leading to a steady gain in the RV precision. Details about the historical changes in the HARPS DRS algorithms  may be found in following series of papers: \cite{1996A&AS..119..373B, 2002A&A...388..632P, 2007A&A...468.1115L, 2009A&A...493..639M, 2009A&A...507..487M}.  

Measurements used in this paper were reprocessed with version 3.5 of the HARPS DRS \footnote{All data used in the analysis will be available via VizieR at CDS.}

A description of the HARPS data used in this study is given in Table~\ref{tab:full_data_stats} while \mystar's stellar parameters are summarised in Table~\ref{tab:hoststar}. \mystar\ has been observed by HARPS $246$ times over a span of more than $10$~years, achieving consistent quality of mean($\sigma_{\rm{RV}}$) = 0.76 \mps.

\begin{table}
\centering
    \caption{\mystar: Stellar properties,  with corresponding references (1) \citet{Gray2006};
    (2) \citet{kharchenko2001};
    (3) \citet{skrutskie2006};
    (4) \citet{Santos2013}; 
    (5) \citet{delgado2019stellardata};
    (6)  GAIA DR2 \citet{GAIADR2}}
    \label{tab:hoststar}
    \begin{tabular}{l c c}
    \hline
    \multicolumn{2}{c}{HD13808} & Reference\\ \hline
        Spectral Type & K2V  & (1)\\
        V (mag) & $8.38 \pm 0.01$ & (2)\\
        K (mag) & $6.25 \pm 0.02$ & (3)\\
        T$_{\rm{eff}}$ (K) & $5035 \pm 50$  & (4)\\
        Fe/H (dex) & $-0.21 \pm 0.02$ & (4)  \\
        $M\ (M_	\odot)$ & $0.771 \pm 0.022$ & (5)\\
        Age (Gyr) & $7.2 \pm 4.8$ & (5) \\
        $R_S\ (R_\odot)$ & $0.781^{+0.017}_{-0.022}$ & (6) \\
        d (pc) & $28.2535 \pm 0.0256$ & (6) \\
    \hline
    \end{tabular}
\end{table}

\section{Modelling Radial Velocities}
\label{chapter:RV}
To model the observed RV of a star at time $t_i$ with $N_p$ planets, we sum the planets' individual Keplerian terms, neglecting planet-planet interactions; following \citet{Feroz2011} and the formalism given in \citet{Balan2009} we arrive at:
\begin{equation}
    {\rm RV}(t_i) = V_i + \sum_{p=1}^{N_p}  K_p \left[  \cos(f_{i,p} + \omega_p) + e_p \cos(\omega_p)\right],
    \label{equ:keplerians}
\end{equation}
where $V_i$ is the systemic velocity, $K_p$ the RV semi-amplitude, $f_{i,p}$, the true anomaly, $e_p$ the orbital eccentricity, and $\omega_p$ the argument of periastron of the $p$\th\ planet, respectively. In addition, the introduction of the mean longitude $\lambda_p$ of the $p$\th\ planet is necessary as part of the computation of $f_{i,p}$ which requires $e_p$ and the period of the planet $P_p$ as well.

A planet's semi-amplitude $K_p$ is related to its mass $M_p$, period and orbital eccentricity $e_p$, as well as to the mass of the parent star $M_S$ and the inclination of the system $i$ relative to the observer:
\begin{equation}
    K_p = \frac{28.4329\ \mps}{\sqrt{1 - e_p^2}} \left( \frac{P_p}{1\ \mathrm{yr}}\right)^{-1/3} \frac{M_p \sin (i)}{M_{\mathrm{Jup}}} \left(\frac{M_S}{M_\odot}\right)^{-2/3}.
\end{equation}

In summary, we have five free parameters per planet ($K$, $\omega$, $e$, $P$, and $\lambda$) plus a white-noise `jitter' term $\sigma^+_{\rm{RV}}$ which is added to the observational error in quadrature (see Section~\ref{sec:likelihood}) and any terms for describing the systemic velocity, $V$ -- typically a constant offset plus a linear or possibly quadratic polynomial term. For the analysis in this paper, a quadratic polynomial term was used for all models. Additionally, our models for describing stellar activity contributions to RVs contain anything from one to sixteen free parameters; these models are described in detail in Section~\ref{chapter:stellar}.

The uninformative priors for our five free parameters per planet are listed in Table~\ref{tab:priors}, as well as the priors for the polynomial terms and for the additive white-noise term. Note that the lower and upper boundary for planet periods was set to $5$ and $100$~d for computational efficiency. This choice was supported, in the first instance, by the fact that the Lomb-Scargle power spectrum of RVs revealed no significant periodicities above about $60$~d (see Fig.~\ref{fig:bgls_activity}). Moreover, models allowing planets with periods longer than $100$~d or shorter than $5$~d always had lower Bayesian evidences than models without such planets (when using a GP stellar activity model -- cf.~Section~\ref{chapter:results_HD13808}), and the inferred RV semi-amplitudes of these short- or long-period planets was always consistent with zero.\footnote{However, the computational cost of evaluating models with expanded planet-period priors was typically an order of magnitude greater than when using the more restricted prior, and posterior multi-modality was far more pronounced.}

The priors used for stellar activity model parameters were generally also chosen to be uninformative, and are discussed in Section~\ref{chapter:stellar} and summarized in Table~\ref{tab:priors_stellar}.

\begin{table}
    \centering
    \caption{Priors for the Keplerian, RV trend and additive white noise (`jitter') parameters used in all of our models and analyses. \newline * We also require for the periods to be sorted i.e.\ $P_b$ < $P_c$ etc. 
    \newline ** We also require that the corresponding eccentricity e < 1. \newline *** This prior applies to all polynomial terms.}
    \label{tab:priors}
    \begin{tabular}{l c c c}
    \hline
    Parameter & Prior & Lower Bound & Upper Bound\\
    \hline
    $P$ (d)* & Log Uniform & $5$ & $100$ \\ 
    $K$ (\mps) &  Log Uniform & $0.1$ & $10$ \\ 
    $\sqrt{e}\ \sin( \omega)$**&  Uniform & $-1$ & $1$\\ 
    $\sqrt{e}\ \cos( \omega)$**&  Uniform & $-1$ & $1$ \\ 
    $\lambda$ (rad)     &  Uniform & $0$ & $2\uppi$\\ 
    $V$ (\mps)***   &  Uniform & $-\rm{RV}_{\max}$ &  $\rm{RV}_{\max}$ \\ 
    $\sigma^+_{\rm RV}$ (\mps) & Uniform & 0 & 20 \\ \hline
    \end{tabular}
\end{table}

\section{Modelling Stellar Activity}
\label{chapter:stellar}
In this section we introduce the multiple ways of modelling stellar activity applied in this work. Simple models like linear dependencies are considered, as well as harmonic modelling of the rotation period of the star, combining BIS and RV measurements in a simultaneous fit. We also discuss more sophisticated approaches to modelling activity-induced RV variations, namely the $FF'$ approach and simultaneous GP regression over multiple activity indicators.

\begin{table}
    \centering
    \caption{Priors for the parameters of the various stellar activity models described in Section \ref{chapter:stellar}. \newline * $R_S$ is the radius of the star and $\sigma_S$ the estimated error. \newline $^+$ $\sigma$ is the observed standard deviation of the FWHM.}
    \label{tab:priors_stellar}
    \begin{tabular}{l c c c}
    \hline
    Parameter & Prior & Lower Bound & Upper Bound\\
    \hline
    $\alpha$ &  Uniform & $-\rm{RV}_{\max}$ &  $+\rm{RV}_{\max}$\\ 
    $P_{\rm{rot}}$ & Uniform & $30$ d & $42$ d \\ 
    $P_{\rm{magn}}$ & Uniform & $500$ d & $4500$ d \\ 
    $\phi$&  Uniform & $0$ & $2 \uppi$\\ 
    $C_{\rm{BIS}}$     &  Uniform & $-\rm{BIS}_{\max}$ &  $+\rm{BIS}_{\max}$ \\
    $C_{\rm{magn}}$     &  Uniform & $-{\ensuremath{\log R'_{\rm HK, max}}}$ &  $+{\ensuremath{\log R'_{\rm HK, max}}}$ \\
    $\beta$     &  Uniform & $-\rm{RV}_{\max}$ &  $+\rm{RV}_{\max}$ \\
    $\Psi_0$ &  Uniform & $\rm{FWHM}_{\max}$ &  $\rm{FWHM}_{\max}+15\sigma^+$ \\
    $\delta V_c \kappa$ &  Uniform & $0$ &  $10\,000$ \\
    $R_*$ &  Uniform & $R_S-5\sigma_S$* &  $R_S+5\sigma_S$* \\
    \hline
    $V_c$ &  Uniform & $-\rm{rms}(\rm{RV})$ &  $+\rm{rms}(\rm{RV})$ \\
    $V_r$ &  Uniform & $-\rm{rms}(\rm{RV})$ &  $+\rm{rms}(\rm{RV})$ \\
    $L_c$ &  Uniform & $-\rm{rms}(\lrhk)$ &  $+\rm{rms}(\lrhk)$ \\
    $B_c$ &  Uniform & $-\rm{rms}(\bis)$ &  $+\rm{rms}(\bis)$ \\
    $B_r$ &  Uniform & $-\rm{rms}(\bis)$ &  $+\rm{rms}(\bis)$ \\
    $P_{\rm{GP}}$ &  Uniform & $10$~d &  $100$~d \\
    $\lambda_{\rm{p}}$ &  Jeffreys & $0.01$ &  $10$ \\
    $\lambda_{\rm{e}}$ &  Jeffreys & $10$~d &  $400$~d \\
    \hline
    \end{tabular}

\end{table}

\subsection{Linear activity model}
\label{sec:linear_modelling}

A linear relation between RVs and BIS was considered as seen in equation \eqref{equ:BIS_linear}, where $\rm{RV}_{{\rm{Kepler}}} (t)$ are the RV signatures of the planets at time $t$ as described previously in equation \eqref{equ:keplerians}; $\rm{BIS} (t)$ represents the BIS measurements taken at time $t$ and $\alpha$ describes the free parameter for the linear relation. A quadratic polynomial term including a constant offset is represented by $V$.

\begin{equation}
{\rm{RV}}_{{\rm{total}}}(t) = {\rm{RV}}_{{\rm{Kepler}}}(t) + \alpha {\rm{BIS}}(t) + V.
    \label{equ:BIS_linear}
\end{equation}

This method has already been successful in identifying false positives \citep[e.g][]{Didier2001}. However, note that it has been shown that there can be a temporal offset between activity and associated RV variations on time scales of few days causing a `blur' of any linear correlation between those two parameters, making a linear model a choice with a major caveat \citep[e.g][]{santos2014, Cameron2019}. The prior distribution for $\alpha$ is shown with the other parameters related to stellar activity modelling in Table~\ref{tab:priors_stellar}.

In a similar way, a linear dependency on the \lrhk\ activity indicator was considered as we hoped to capture any possible linear long term relationship with
\begin{equation}
{\rm{RV}}_{{\rm{total}}}(t) = {\rm{RV}}_{{\rm{Kepler}}}(t){\rm{ + }}\beta \lrhk (t) + V
    \label{equ:logrhk_linear}
\end{equation}
where $\beta$ represents the linear factor and $\lrhk (t)$ the \lrhk measurements at time $t$. The prior distribution for $\beta$ can be found in Table~\ref{tab:priors_stellar}. 

\subsection{Harmonic activity model}
\label{sec:harmonic}
Our second activity model entailed fitting a sinusoid to both the RVs and BIS time series, enforcing an identical period but allowing different phases and amplitudes between the two. We assume that this captures solar spots and other stellar features which are sensitive to the rotation of the star. In addition, it accounts for a likely time shift which is a problem when considering linear correlations as mentioned in Section~\ref{sec:linear_modelling}.

Equation \eqref{equ:RV_rotation_P} and \eqref{equ:BIS_rotation_P} show the two models with the harmonics of the rotation period simultaneously fit to the RV and BIS measurements, respectively. The parameter $P_{\rm{rot}}$ is the putative rotation period of the star fit to both models, while the other parameters $K_i, \phi_{\mathrm{RV}}$ and $K_k,\phi_{\mathrm{BIS}}$ and $C_{\mathrm{BIS}}$ describe the sinusoidal signal of the $i$\th \ and $k$\th \ harmonic for the $\mathrm{RV}(t)$ and $\mathrm{BIS}(t)$ model respectively, with $N_h$ being the number of harmonics. As before, the parameter $V$ represents a quadratic polynomial contribution.

\begin{equation}
{\mathrm{RV}}_{\mathrm{total}}(t) = {\mathrm{RV}}_{\mathrm{Kepler}}(t) + \sum\limits_{i = 1}^{{N_h}} {{K_i}\sin \left( {\frac{{2\uppi it}}{P_{\mathrm{rot}}} + \phi _{\mathrm{RV}}} \right)} + V 
    \label{equ:RV_rotation_P}
\end{equation}
\begin{equation}
{\mathrm{BIS}}(t) = \sum\limits_{k = 1}^{N_h} {K_k}\sin \left( \frac{{2\uppi kt}}{P_{{\mathrm{rot}}}} + \phi _{\mathrm{BIS}} \right) + C_{\mathrm{BIS}}
    \label{equ:BIS_rotation_P}
\end{equation}

 The prior distribution for $K$ is as in Table~\ref{tab:priors}; the priors for $\phi, C_{\rm{BIS}}$ and $P_{\rm{rot}}$ are displayed in Table~\ref{tab:priors_stellar}. Note that the prior on the rotation period $P_{\rm{rot}}$ was chosen to cover a narrow range based on the following: (i) the rotation period for \mystar \ was estimated before to be $\sim 40$~d based on the average of \lrhk \ measurements by \citet{lovis2011}; (ii) signals likely corresponding to the harmonics of the rotation period are detected in the periodograms of the stellar activity indicators, see Section~\ref{sec:periodogram} (Fig.~\ref{fig:bgls_activity}). Further motivation was provided by the fact that (iii) when considering zero or 1-planet models, the rotation period of the star became locked on to the period of one of the two planets, and that (iv) during preliminary runs, the MAP value for the rotation period never went above $40$~d. By limiting the period to below $42$~d we reduced the computational burden of our modelling.
 
 \subsection{Long-term magnetic activity model}
\label{sec:magnetic}
In a similar fashion to the harmonic activity model, the `magn. cycle' model fits a sinusoid to two data sets, this time to the RVs and the \lrhk \ time series, enforcing an identical period but allowing different phases and amplitudes. We assume that this captures stellar activity corresponding to a long-term magnetic activity cycle.

Equation \eqref{equ:RV_magn} and \eqref{equ:BIS_magn} show the two models with the cycle period $P_{\mathrm{magn}}$ simultaneously fit to the RV and \lrhk\ measurements, respectively. The parameters $K_{\mathrm{magn,RV}}$, $\phi_{\mathrm{magn,RV}}$ and $K_{\mathrm{magn}}$,$\phi_{\mathrm{magn}}$ and $C_{\mathrm{magn}}$ describe the $\mathrm{RV}(t)$ and $\mathrm{\lrhk}(t)$ model respectively; $V$ represents a constant offset and a quadratic polynomial term.

\begin{equation}
{\mathrm{RV}}_{\mathrm{total}}(t) = {\mathrm{RV}}_{\mathrm{Kepler}}(t) + {{K_{\mathrm{magn,RV}}}\sin \left( {\frac{{2\uppi t}}{P_{\mathrm{magn}}} + \phi _{\mathrm{magn,RV}}} \right)} + V 
    \label{equ:RV_magn}
\end{equation}
\begin{equation}
{\mathrm{\lrhk}}(t) = {K_{\mathrm{magn}}}\sin \left( \frac{{2\uppi t}}{P_{{\mathrm{magn}}}} + \phi _{\mathrm{magn}} \right) + C_{\mathrm{magn}}
    \label{equ:BIS_magn}
\end{equation}

 The prior distribution for $K$ is as in Table~\ref{tab:priors}; the priors for $\phi, C_{\rm{magn}}$ and $P_{\rm{magn}}$ are displayed in Table~\ref{tab:priors_stellar}. The upper and lower prior limits for $P_{\rm{magn}}$ were determined by a broad peak around $1000$---$4000$~d in the periodogram of \lrhk as well as by the obvious periodic signal in the \lrhk \ measurements (see Fig.\ \ref{fig:HD13808_RHK}).

\subsection{$FF'$ method}
\label{sec:ffprime}

The FWHM measurements were used for applying the $FF'$ method which was introduced by \citet{Aigrain2012} as a method for relating the photometric brightness and RV variations of a star. It uses the flux of the star $\Psi (t)$ and its derivative $\Dot{\Psi} (t)$ as an indicator for spot coverage and predicts RV variations. These variations include the RV perturbation $\Delta \rm{RV}_{\rm{\rm{rot}}} (t)$ due to the presence of spots on the rotating photosphere and their effect of the suppression of convective blueshift $\Delta \rm{RV}_{\rm{conv}} (t)$. 

\citet{Aigrain2012} describes these with the following two equations where $f$ represents the drop in flux produced by a spot at the centre of the stellar disc, $R_*$ is the stellar radius, $\delta V_c$ is the difference between the convective blueshift in the unspotted photosphere and that within the magnetized area and $\kappa$ is the ratio of this area to the spot surface.
\begin{equation}
    \Delta \rm{RV}_{\rm{\rm{rot}}} (t) = -  \frac{\Dot{\Psi} (t)}{\Psi_0} \left[ 1 - \frac{\Psi (t)}{\Psi_0}  \right] \frac{R_*}{f}
\end{equation}
\begin{equation}
    \Delta \rm{RV}_{\rm{conv}} (t) = \left[ 1 - \frac{\Psi (t)}{\Psi_0} \right]^2 \frac{\delta V_c \kappa}{f}
\end{equation}

The total RV variation $\Delta \rm{RV}_{\rm{activity}}$ created by stellar activity is then the sum of both terms:
\begin{equation}
    \Delta \rm{RV}_{\rm{activity}} = \Delta \rm{RV}_{\rm{\rm{rot}}} (t) + \Delta \rm{RV}_{\rm{conv}} (t)
\end{equation}

The parameter $f$ is approximated as
\begin{equation}
    f \approx \frac{\Psi_0 - \Phi_{\rm{min}}}{\Psi_0}
\end{equation}
with $\Phi_{\rm{min}}$ being the minimum observed flux.

It has been argued that CCF FWHM and \lrhk measurements should both behave, to first order, as the convective blueshift suppression term in the $FF'$ method, which in turn is a close proxy for the integrated active region coverage on the visible stellar hemisphere \citep{RajpaulGP}; indeed, it has been shown in practice that the CCF FWHM can be a good tracer of photometric flux \citep[e.g.\ ][]{SuarezMascareno2020Proxima}. Thus, as we did not have photometric measurements of \mystar\ available, we chose to interpret the FWHM as a proxy for the flux $\Psi (t)$. The derivative of the `flux' was computed numerically by modelling the FWHM with a GP using the \textsc{celerite} algorithm by \citet{celerite}. In total, there were three free parameters in this stellar activity model, $\delta V_c \kappa$, $R_*$ and $\Psi_0$ - their respective priors appear in Table~\ref{tab:priors_stellar}.

\subsection{GP regression model}\label{sec:GP-model}

We used the GP framework developed by \citet{RajpaulGP}, hereafter R15, to model RVs simultaneously with \lrhk\ and \bis\ observations. In short, this framework assumes that all observed stellar activity signals are generated by some underlying latent function $G(t)$ and its derivatives; this function, which is not observed directly, is modelled with a Gaussian process \citep{rasmussen2006,roberts2013}. 

Following R15, activity variability in the RV, \lrhk\ and \bis\ time series can be modelled as:
\begin{align}
\Delta \rm{RV} &= V_c G(t) + V_r \dot{G}(t), \\
\lrhk &= L_c G(t), \rm{ and} \\
\bis &= B_c G(t) + B_r \dot{G}(t),
\end{align}
respectively. The coefficients $V_c$, $V_r$, $L_c$, $B_c$ and $B_r$ are free parameters relating the individual observations to the unobserved Gaussian process $G(t)$. In R15's framework, $G(t)$ itself can be loosely interpreted as representing the projected area of the visible stellar disc covered in active regions at a given time; the GP describing $G(t)$ is assumed to have zero mean and covariance matrix $\mathbf{K}$, where $K_{ij} = \gamma(t_i,t_j)$. As in R15, we adopt the following quasi-periodic covariance kernel function:

\begin{equation}
\gamma ({t_i},{t_j}) = \exp \left[ { - \frac{{{{\sin }^2}\left[ {\uppi ({t_i} - {t_j})/{P_{{\rm{GP}}}}} \right]}}{{2\lambda _{\rm{p}}^2}} - \frac{{{{({t_i} - {t_j})}^2}}}{{2\lambda _{\rm{e}}^2}}} \right],
\end{equation}
where $P_{\rm{GP}}$ is the period of the quasi-periodic activity signal, $\lambda_{\rm{p}}$ is the inverse harmonic complexity of the signal (such that signals become sinusoidal for large values of $\lambda_{\rm{p}}$, and show increasing complexity/harmonic content for small values of $\lambda_{\rm{p}}$), and $\lambda_{\rm{e}}$ is the time scale over which activity signals evolve. This quasi-periodic covariance kernel has been widely used to model stellar activity signals in both photometry and RVs \citep[e.g.][]{haywood2014disentangling,RajpaulGP,grunblatt2015determining,bonfils2018temperate}. The full expressions for the covariance between the three different observables modelled are given in R15.

We expect, in principle at least, that this sophisticated GP-based approach to modelling RVs jointly with activity indicators should enable more reliable planet characterization, for several reasons. Firstly, by modelling multiple activity-sensitive time series simultaneously (e.g.\ \lrhk, BIS, FWHM, etc., or some subset of these), more information can be gleaned on activity signals in RVs, compared to exploiting only simple correlations between RVs and (typically) one of these time series. Additionally, the framework uses GP draws and derivatives thereof as basis functions for modelling available time series, rather than e.g.\ sinusoids or other simple parametric models, the inappropriate use of which could easily lead to the introduction of correlated signals into model residuals. The GP basis functions could in principle take any form, although in the GP framework their properties are constrained to some extent by the data itself, and from reasonable prior assumptions about the quasi-periodic nature of stellar activity signals. The GP framework also incorporates the $FF'$ formalism directly as a special case; the former approach may be thought of as a generalization of the latter.

\section{Bayesian Inference and \textsc{PolyChord}}
\label{chapter:Bayesian}
\subsection{Bayesian model comparison}
As we shall use Bayesian inference to evaluate the relative posterior probabilities of different models, we summarise briefly here the relevant formalism. Firstly, Bayes' Theorem, given in equation \eqref{equ:Bayes}, is used to relate (i) the posterior probability $\Pr(\Theta | D, M) = \mathcal{P}(\Theta)$ of the parameters $\Theta$ given data $D$ and a model $M$ to (ii) the prior distribution $\Pr(\Theta | M) = \pi(\Theta)$ of $\Theta$ given $M$, (iii) the likelihood $\Pr(D | \Theta, M)=\mathcal{L}(\Theta)$ of $D$ given $\Theta$ and $M$ and the Bayesian evidence $\Pr(D|M) = \mathcal{Z}$ of $D$ given $M$. Following the notation used by \citet{Feroz2009}:
\begin{equation}
    \Pr(\Theta | D, M) = \frac{\Pr(D | \Theta, M)\ \Pr (\Theta | M)}{\Pr(D|M)},
    \label{equ:Bayes}
\end{equation}
or simply
\begin{equation}
    \mathcal{P}(\Theta)= \frac{\mathcal{L}(\Theta)\ \pi(\Theta)}{\mathcal{Z}}.
    \label{equ:bayes_post}
\end{equation}

For model selection, one can compare two models $M_1$ and $M_2$, given data $D$, by computing the ratio of their respective posterior probabilities; this ratio is also known as the Bayes factor $R$: 
\begin{equation}
    \mathcal{R} = \frac{\Pr(M_1 | D)}{\Pr(M_2 | D)} = \frac{\Pr(D | M_1)\ \Pr(M_1)}{\Pr(D | M_2)\ \Pr(M_2)} = \frac{\mathcal{Z}_1 }{\mathcal{Z}_2 } \frac{\Pr(M_1) }{\Pr(M_2)}. 
    \label{equ:post_distr_comparison}
\end{equation}
Note that $\frac{\Pr(M_1) }{\Pr(M_2)} $ is the relative a priori probability between the two models, which is usually set to one.

To decide whether the evidence difference is significant to favor one model over the other, we make use of the Jeffreys scale as given in Table \ref{tab:jeffreys}, where the logarithmic scale of the evidence difference is used. This simplifies equation \eqref{equ:post_distr_comparison} to:
\begin{equation}
    \ln (\mathcal{R}) = \ln \left( \frac{\mathcal{Z}_1}{\mathcal{Z}_2} \right)= \ln \mathcal{Z}_1- \ln \mathcal{Z}_2
\end{equation}

\begin{table}
    \centering
    \caption{Jeffreys' scale, as introduced by \citet{jeffreys1983theory}, for interpreting differences in Bayesian evidences. This scale interprets the strength of evidence favouring one model over another. }
    \label{tab:jeffreys}
    \begin{tabular}{c c c c}
    \hline
        |$\ln \mathcal{R}$| & Odds & Probability & Remark  \\ \hline
        < 1.0 & $\lesssim$  3:1 & < 0.750 & Inconclusive \\
        1.0 &  $\sim$ 3:1 &  0.750 & Weak evidence \\
        2.5 & $\sim$12:1 & 0.923 & Moderate evidence \\
        5.0 & $\sim$150:1 &  0.993 & Strong evidence \\ \hline
    \end{tabular}
\end{table}

\subsection{Likelihood function}
\label{sec:likelihood}
It is commonly assumed that an additive white Gaussian noise (AWGN) model is sufficient for describing observational noise, as in probability theory the central limit theorem states that the sum of independent random variables tends towards a Gaussian distribution, even if the original ones are not normally distributed \citep{fischer2011}. In this case, the likelihood $\mathcal{L}(\Theta)$ of parameters $\Theta$ can be written as: 
\begin{equation}
    \mathcal{L}(\Theta) = \prod_{i=1}^N \frac{1}{\sqrt{2\uppi \sigma_i^2}}\ \exp\left( -\frac{[v(t_i; \Theta) - v_i]^2}{2 \sigma_i^2}\right), 
    \label{equ:likelihood}
\end{equation}
where $v(t_i; \Theta)$ describes the model's predicted RV for parameters $\Theta$, while $v_i$ describes the RV observed at time $t_i$, with corresponding error estimate $\sigma_i$ which contains the observational error and the additive noise `jitter' term $\sigma^+_{\rm{RV}}$ added in quadrature.  
The logarithmic likelihood, which is more convenient to work with, is computed as:
\begin{equation}
    \ln \mathcal{L}(\Theta) = \sum_{i=1}^N  -\ln\sqrt{2\uppi \sigma_i^2} - \frac{1}{2 \sigma_i^2}[v(t_i; \Theta) - v_i]^2.
    \label{equ:loglikelihood}
\end{equation}
Note that the above formalism applies to all of our models \emph{except} for the GP model, which explicitly generalises the AWGN model by allowing for red (correlated) noise. In this case, the log likelihood may be computed via the more general expression
\begin{equation}
\ln {{\cal L}(\Theta)} = - {\textstyle{N \over 2}}\ln 2\upi - {\textstyle{1 \over 2}}\ln \det {{\bf{K}}} - {\textstyle{1 \over 2}}{{\bf{r}}}^{\rm{T}}{{\bf{K}}}^{ - 1}{{\bf{r}}}  ,
\label{eq:NLL_GP}
\end{equation}
where $\mathbf{K}$ is the matrix defining the covariance between all pairs of observations (see R15), and $\mathbf{r}$ is a vector of residuals with the $i$\th\ element given by $v(t_i; \Theta) - v_i$. Note that in the special case where $\mathbf{K}$ is a diagonal matrix with the $i$\th\ element given by $\sigma_i^2$, i.e.\ where the noise is assumed to be white, equation \ref{eq:NLL_GP} reduces to equation \ref{equ:loglikelihood}. 

As the Keplerian model requires the Kepler equation to be solved, we made use of the efficient CORDIC-like method introduced by \citet{Zechmeister2018} where double precision is obtained within 55 iterations.

\subsubsection{Avoiding unstable Keplerian orbits}
While computing likelihoods we checked if each pair of planets would be stable following the criterion introduced by \citet{stabilityGladman}: $\Delta > 2 \sqrt{3}\ R_H(i,j)$ where $\Delta = a_j - a_i$ is the difference between the semi-major axis of the $i$-th and $j$-th planet and $R_H(i,j)$ the planets' mutual Hill radius. This has been used before e.g.\ by \citet{Malavolta2017}. 

If the parameters drawn resulted in a planet system which was considered unstable, the likelihood was set to zero, i.e.\ $Z_0=0$; in our case, as we worked with log likelihoods, this was approximated numerically by $\ln{Z_0}=-1\times10^{30}$.

\subsection{\textsc{PolyChord}}
We used \textsc{PolyChord} as it is an state-of-the-art nested sampling algorithm, designed to work with very high dimensional parameter spaces \citep{PolyChord}. A short discussion comparing it with MultiNest \citep{Feroz2009} can be found in \citet{Richard2018}. It was developed using C++ and Fortran and can also be called as a Python package. 

In order to test the reliability of our algorithm with \textsc{PolyChord} we used two sets of simulated data. One data set contained random uncorrelated Gaussian noise (with a standard deviation of $1 \mps$) without any planetary signals, while the other one consisted of two planets on elliptical orbits with the same Gaussian noise. Both data sets also included an offset of a few \mps\ and linear trend of $10^{-5}~\mps$ per day. Models of $0$--$3$ planet signals with long term trends in the form of second-order polynomials were fitted and their evidences compared. Both elliptical and circular orbital solutions were computed. We note here that in general, the evidence uncertainty reported by \textsc{PolyChord} is an underestimate, and it is better practice to estimate it via the range of scatter in the evidence across multiple runs \citep{higson2018,PlanetEvidence}; we used the latter approach throughout our analyses. (It is probable that this behaviour would also be ameliorated by increasing the number of live points beyond the computational resources available at the time of writing.)

For the first data set without planetary signals, the algorithm successfully favoured the no-planet model consisting of a second-order polynomial. Every other model computed showed lower evidences, with at least $\ln\mathcal{R} \approx 2$ for the zero-planet model vs.\ one sinusoidal signal or $\ln\mathcal{R} \approx 3$ for the zero-planet model vs.\ one elliptic orbital signal; evidences decreased with the number of signals fitted. 

Similarly, the two planets with eccentric orbits in the second test were also recovered: the preferred model featured two planets, with the second best model containing three planets with eccentric orbits, albeit with a significantly lower evidence than the two-planet model, and thus rejected. The typical rms of the log evidences was of order $\sim3$ for circular orbital fits and of order $\sim 1$ for eccentric orbital fits.

\begin{figure}
    \centering
    \includegraphics[width=\columnwidth]{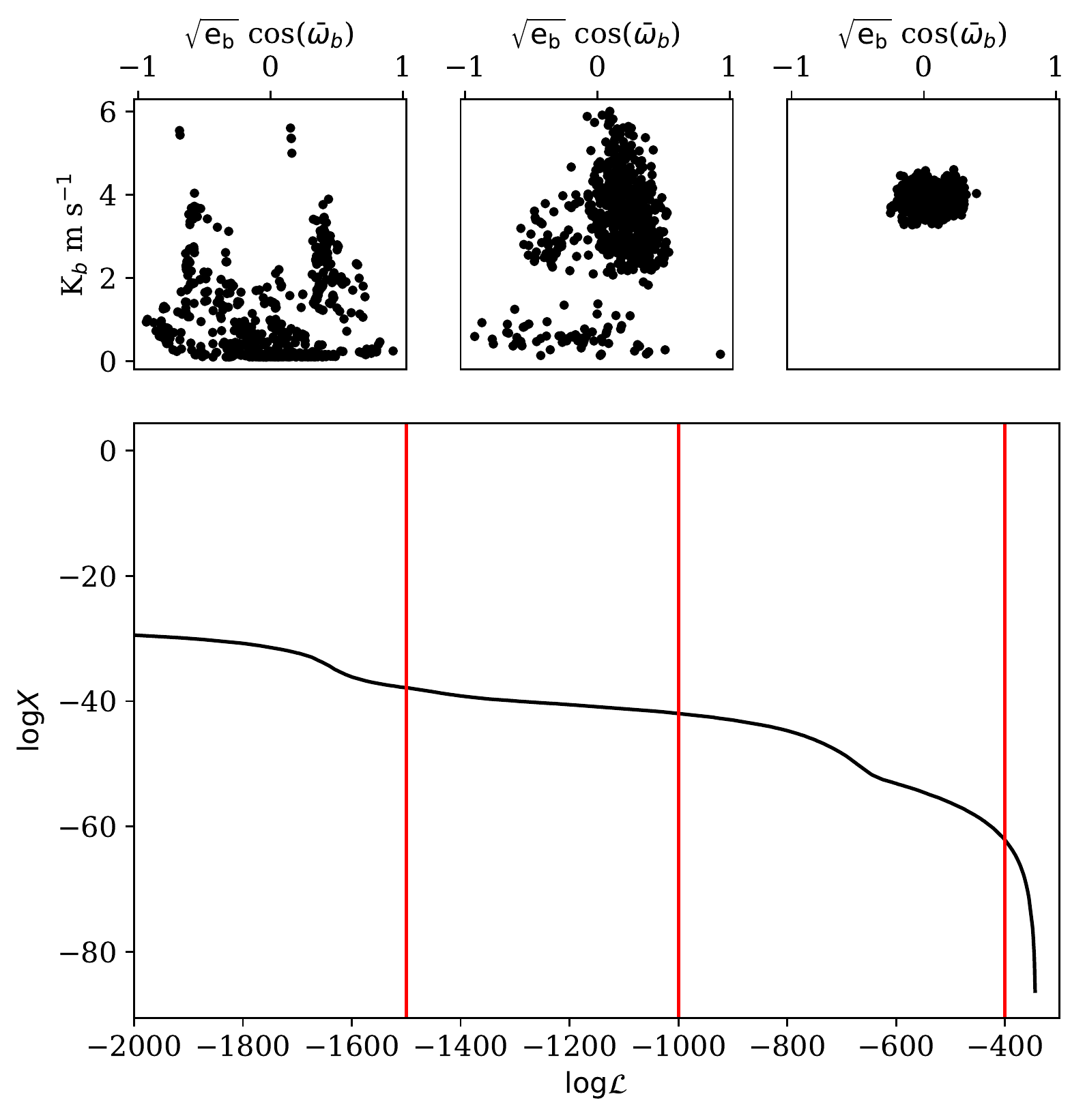}
    \caption{A typical run with \textsc{PolyChord}. The bottom graph shows the logarithmic likelihood over the course of the run, with $\log X$ corresponding to the prior volume i.e.\ at the start of the run $\log X$ is at its highest value. Two phase transitions are clearly visible as 'knees' in the logL-logX curve. On the top, the live-point distributions of two parameters ($K_b$ and $\sqrt{e_b}\ \cos( \omega_b)$) within a given likelihood contour (indicated by the red vertical lines) are shown, demonstrating the transitions between different phases. Note that if the precision criterion is not set low enough, one runs the risk of stopping nested sampling at too early a $\log X$ value (i.e.\ at one of the lower knees), and this is indeed what we observed in preliminary tests.}
    \label{fig:polychord_phase}
\end{figure}

Over multiple runs with complex models, our tests showed that the default stopping criterion of \textsc{PolyChord} was not precise enough\footnote{The convergence criterion in \textsc{PolyChord} is defined as the point where the fraction of total evidence contained in the live points drops below the default value of $10^{-3}$ \citep{PolyChord}.} for these sort of exoplanet applications. This resulted in inconsistencies where the sampling in some runs missed transitions to areas of higher likelihood (see Fig.~\ref{fig:polychord_phase}) and showed very different evidence values and posterior distributions. To ensure consistent and robust runs, the convergence criterion was lowered to values between $10^{-5}$ to $10^{-12}$ for runs with a large number of planets and complex stellar activity models. As an historical aside, nested sampling was in fact invented \citep{skilling2006} in order to solve precisely these kind of phase transition problems.

\section{Analysis and Results}
\label{chapter:results_HD13808}
\begin{table}
    \centering
     \caption{An overview of the HARPS RV data subsets of \mystar\ referred to as low and high activity when \lrhk is $< - 4.90$ and $> - 4.90$, respectively.}
    \label{tab:data_stats}
    \begin{tabular}{lcc}
    \hline
         
          & \multicolumn{1}{c}{Low Activity} & \multicolumn{1}{c}{High Activity}\\ \hline
         \lrhk & $<-4.90$ & > $-4.90$ \\
         N data & 123  & 123\\
          $\Delta$T (d) &  3332 &  3701\\
         rms(RV) (\mps) & 3.53 & 3.78\\
         mean($ \sigma_{\rm{RV}}$) (\mps)& 0.72 & 0.78\\
          mean(\lrhk) & $-4.95$ & $-4.83$ \\
         \hline  
    \end{tabular}

\end{table}

\begin{figure}
    \centering
    \includegraphics[width=0.46\textwidth]{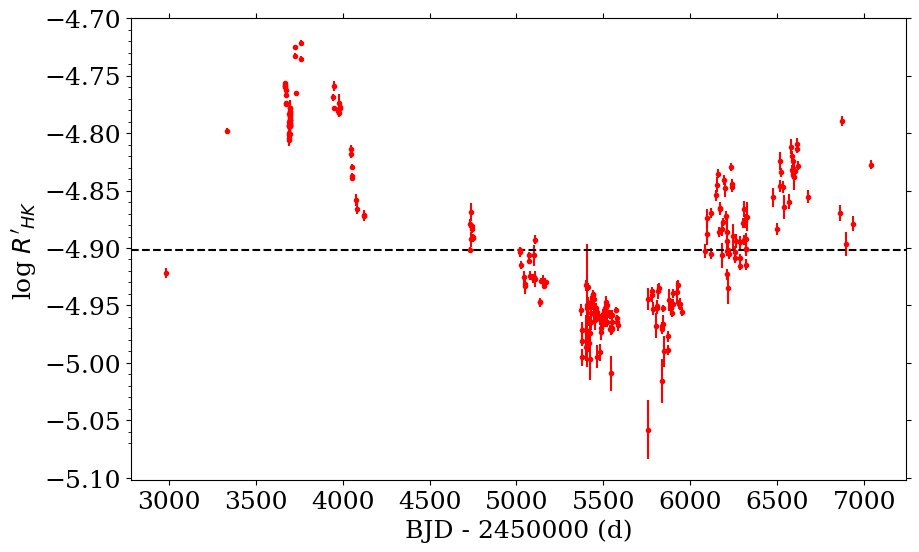}
    \caption{Measurement of the $\lrhk $ activity index on the series of spectra of HD~13808. The black dashed line represents the median value of $-4.90$ which was used as the condition for separating the data set into two parts.}
    \label{fig:HD13808_RHK}
\end{figure}

The extent of stellar activity observed from \mystar\ was investigated by looking at the activity indicators BIS, FWHM and \lrhk. None of them showed an obvious linear correlation with the RV measurements. Note that this can occur e.g.\ when there \emph{is} a relation but with a simple time lag, as demonstrated with the case of the Sun \citet{Cameron2019}. However, the \lrhk\ measurements do show a notable cycle from a low of $-5.10$ to a high of $-4.70$, as displayed in Fig.~\ref{fig:HD13808_RHK}. A cycle with similar shape and period is also seen in the BIS and FWHM time series, although with smaller amplitude. We thus decided to extend our analysis by splitting the full RV data into two subsets based on the median(\lrhk) = $-4.90$; low- and high-activity subsets then corresponded to observations with $\lrhk < -4.90$ and $\lrhk > -4.90$, respectively. Table~\ref{tab:data_stats} summarises the statistics of the full data set and those two subsets. Note that although we call one subset `high activity,' \mystar\ is still classified as an inactive star, as an `active' classification would usually require an average of \lrhk $> -4.75$ \citep{Henry1996}. This \lrhk\  cycle shows great similarities to the one of the Sun with its recent maxima and minima being at \lrhk $\simeq -4.83$ and \lrhk $\simeq -4.96$; the extrapolated \lrhk\  value for the Maunder minimum period is \lrhk $\simeq -5.10$ \citep{Mamajek2008}.

\subsection{Periodogram Analysis}
\label{sec:periodogram}

\begin{figure*}
    \centering
    \includegraphics[width = .9\textwidth]{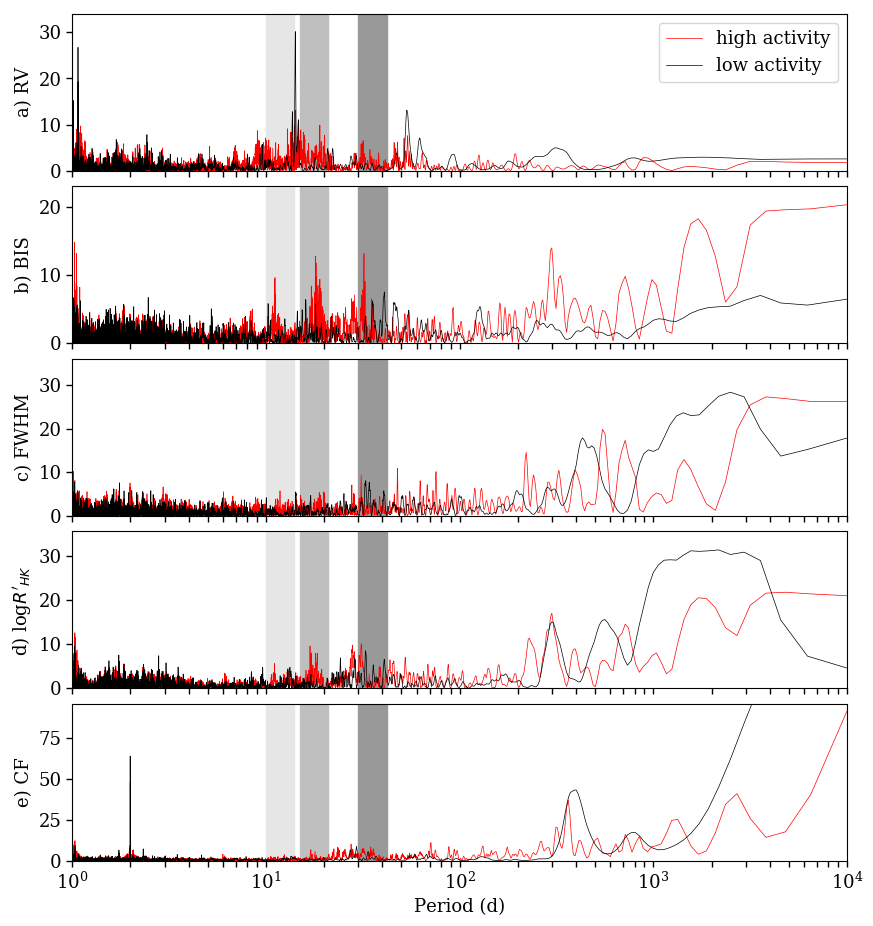}
    \caption{BGLS periodograms of, from top to bottom, the RV, BIS, FWHM and \lrhk  measurements and of a constant function (CF) i.e.\ a dataset with the same timestamps and RV errors but assuming constant RV. The colours red and black indicate the high activity and low activity phases. The dark grey shaded box represents the range of the putative rotation period with its second (grey) and third harmonic (light grey).}
    \label{fig:bgls_activity}
\end{figure*}

To identify strong periodicities in our time series we computed the Bayesian generalized Lomb-Scargle (BGLS) periodogram \citep{Lomb1976,Scargle,GLSP2009,BGLS2015} of the two separated data subsets for the RV, the stellar activity indicators BIS, FWHM and \lrhk and a constant function (CF) with identical time stamps.

The resulting BGLS periodograms are shown in Fig.~\ref{fig:bgls_activity} where peaks in the RV periodograms are visible at $\sim 15$ and $\sim 55$~d. Sets of strong peaks are visible in the BGLS periodograms of the activity indicators \lrhk, BIS and FWHM between $30-40$~d and a few between $25-30$~d.  Strong peaks are evident at $\sim 19$ and $\sim 32$~d in the $\lrhk$ high activity periodogram, with peaks at $\sim 12$, $\sim 19$ and $\sim 34$~d in the BIS periodogram. The FWHM periodograms show peaks in the region of $\sim 35$~d.

The BGLS periodograms of the high- and low-activity subsets of the data suggest that the RV measurements are affected by stellar activity: note, in particular, the excess power around the putative stellar rotation period and the first two harmonics in the high-activity subsets of the BIS and \lrhk\ time series. It is also evident that the stellar activity indicators differ quite strongly when the star is slightly more active; the difference between the mean \lrhk\ value of the two subsets is $\Delta \left<\lrhk\right> = 0.12$.

\subsection{Model Comparison}

\begin{table*}
\centering
\caption{\mystar: Overview of the models used for the study of the \mystar. }
\label{tab:model_overview}
\begin{tabular}{llll}
\hline 
\multicolumn{2}{l}{\rm{Model name}} & {\rm{Description}} \\\hline \hline
\multicolumn{2}{l}{\rm{Circular}}  & \multicolumn{2}{l}{Circular orbit modelling} \\ \hline
\multicolumn{2}{l}{\rm{Kepler}}  &  \multicolumn{2}{l}{Eccentric orbit modelling} \\\hline
&\rm{ + magn. cycle} & \multicolumn{2}{l}{Kepler + long-term magnetic activity cycle; Section~\ref{sec:magnetic}}\\\hline
&\rm{ + linear BIS} & \multicolumn{2}{l}{Kepler + linear dependency on the BIS; Section~\ref{sec:linear_modelling}, equation~\eqref{equ:BIS_linear}}\\\hline
&\rm{ + BIS $P_{\rm{rot}}$ $1^{\rm{st}}$ harm.}& \multicolumn{2}{l}{Kepler + simultaneous fit of $1^{\rm{st}}$ harmonic of $P_{\rm{rot}}$ to the BIS and RVs; Section~\ref{sec:harmonic}} \\\hline
&\rm{ + BIS $P_{\rm{rot}}$ $2^{\rm{nd}}$ harm.} &  \multicolumn{2}{l}{Kepler + simultaneous fit of $1^{\rm{st}}$ and $2^{\rm{nd}}$ harmonic of $P_{\rm{rot}}$ to the BIS and RVs; Section~\ref{sec:harmonic}}\\\hline
&\rm{ + BIS $P_{\rm{rot}}$ $3^{\rm{rd}}$ harm.}& \multicolumn{2}{l}{Kepler + simultaneous fit of $1^{\rm{st}}$, $2^{\rm{nd}}$ and $3^{\rm{rd}}$ harmonic of $P_{\rm{rot}}$ to the BIS and RVs;  Section~\ref{sec:harmonic}}\\\hline
&\rm{ + linear \lrhk} & \multicolumn{2}{l}{Kepler + linear dependency on the \lrhk; Section~\ref{sec:linear_modelling}, equation~\eqref{equ:logrhk_linear}}  \\\hline
&\rm{ + $FF'$ }& \multicolumn{2}{l}{Kepler + FF' method with FWHM as a proxy for flux; Section~\ref{sec:ffprime}} \\ \hline
\multicolumn{2}{l}{\rm{Gaussian process}} & \multicolumn{2}{l}{Simultaneous GP modelling of RVs and activity indicators; Keplerian terms for RVs; Section~\ref{sec:GP-model}}\\\hline\hline
\end{tabular}
\end{table*} 

The different models we considered are summarised in Table~\ref{tab:model_overview}, along with a short description of each. The model including only circular orbital solutions is called `Circular', while the elliptical orbital solutions are referred to as `Kepler'. A `+' indicates that the `Kepler' solution was complemented with a specific stellar activity model, such as linear dependency on the BIS (`Kepler + linear BIS'). This was done equivalently for a subset of models applied to the low and high-activity subsets of the data.

\subsubsection{Results}

\begin{table*}
\caption{\mystar: Relative Bayesian evidences for different number of planets for each model and their scatter -- as well as the residual RV (O$-$C) rms and median absolute deviation (MAD) -- for the 2-planet model (or 1-planet model, if favoured) computed using the respective MAP values (see Table~\ref{tab:model_comparison_parameters}). The model with two planetary signals for each model type was set to be zero. The errors correspond to the standard error on the mean evidence across multiple runs or to the highest individual run error provided by \textsc{PolyChord}, whichever of the two is greater. }
\label{tab:model_comparison_evidence}
\begin{tabular}{llcccccccccc}
\hline
\multirow{2}{*}{\rm{Model}} & & \multicolumn{1}{c}{\rm{No planets}} & \multicolumn{1}{c}{\rm{1 planet}} & \multicolumn{1}{c}{\rm{2 planets}} & \multicolumn{1}{c}{\rm{3 planets}} & \multicolumn{1}{c}{\rm{4 planets}}& & RV residual & RV residual \\ 
& & $\ln (\mathcal{R})$  & $\ln (\mathcal{R})$  & $\ln (\mathcal{R})$  & $\ln (\mathcal{R})$  & $\ln (\mathcal{R})$  & & rms \mps & MAD \mps \\ \hline \hline
\multicolumn{2}{l}{\rm{Circular}}& $-65 \pm 2$  & $-15 \pm 2 $  & $0 \pm 3$  & $+2 \pm 3$& $+6\pm 2$ & & 4.09 & 2.79 \\ \hline
\multicolumn{1}{l}{\rm{Kepler}} & & $-56 \pm 3$  & $-6 \pm 4 $  & $0 \pm 4 $  & $+6 \pm 4 $& $+6\pm 4$  & & 3.23 & 2.31 \\\hline        
& \rm{ + magn. cycle}     & $-78 \pm 3  $  & $-21 \pm 3 $  & $0\pm 3 $  & $+10 \pm 4 $          & $-8 \pm 13$            & & 3.89 &2.41         \\\hline
&\rm{ + linear BIS}     & $-62 \pm 2 $  & $-12 \pm 2 $  & $0 \pm 3$  & $-2 \pm 2 $          &  $0\pm 2$      & & 2.97 & 2.11            \\\hline
&\rm{ + BIS $P_{\rm{rot}}$ $1^{\rm{st}}$ harm.}    & $-51 \pm 4 $  & $-7 \pm 4 $  & $0 \pm 4 $  & $-2 \pm 8 $          & $-10 \pm 6$     & & 2.62 & 1.76                 \\\hline
&\rm{ + BIS $P_{\rm{rot}}$ $2^{\rm{nd}}$ harm.}    & $-52 \pm 2  $  & $-7 \pm 3 $  & $0 \pm 3$  & $-3 \pm 6 $          & $-6 \pm 4 $    & & 2.92 & 1.99                 \\\hline
&\rm{ + BIS $P_{\rm{rot}}$ $3^{\rm{rd}}$ harm.}    & $-71 \pm 4 $  & $-28 \pm 4 $  & $0 \pm 3 $  & $-9 \pm 4 $          & $-15 \pm 4$            & & 2.96 & 1.73         \\\hline
&\hspace{0.7cm} \rm{ + magn. cycle}     & $-61 \pm 6  $  & $-7 \pm 6 $  & $0 \pm 6$  & $-7 \pm 8 $          & $+2 \pm 9$            & & 2.62 & 1.66         \\\hline
&\rm{ + linear \lrhk}     & $-71 \pm 3 $  & $-16 \pm 3 $  & $0 \pm 3 $  & $-6 \pm 4 $ & $-2 \pm 4$           & & 2.18 & 1.51           \\\hline
&\rm{ + $FF'$ } & $-64 \pm 1 $  & $-14 \pm 2 $  & $0 \pm 1 $  & $-3 \pm 5 $          &  $-9 \pm 6$   & & 2.47  & 1.53   \\ \hline 
\multicolumn{2}{l}{\rm{Gaussian process }} & $-73 \pm 1 $   & $-19 \pm 2 $  & $0 \pm 1 $  & $-4 \pm 3 $  & $-23 \pm 6 $  & & 2.18 & 1.45\\ \hline \hline
\multicolumn{2}{l}{\rm{Circular low activity }} & $-45 \pm 3 $   & $-10 \pm 4 $  & $0 \pm 4 $  & $-4 \pm 4 $  & $+9 \pm 4 $  & & 3.08 & 2.08\\\hline
\multicolumn{2}{l}{\rm{Kepler low activity }} & $-37 \pm 1 $   & $-5 \pm 1 $  & $0 \pm 2 $  & $+14 \pm 1 $  & $+7 \pm 2$ & & 2.17 & 1.33 \\\hline
&\rm{ + BIS $P_{\rm{rot}}$ $3^{\rm{rd}}$ harm.}      & $-38 \pm 1 $  & $-16 \pm 2 $  & $0 \pm 2 $  & $+1 \pm 4 $          &  $+3 \pm 2$               & &  2.30 & 1.46   \\\hline
&\rm{ + $FF'$ } & $-38 \pm 4 $  & $-7 \pm 4 $  & $0 \pm 2 $  & $-5 \pm 4 $          &  $+5 \pm 7 $     & &  4.15 & 3.08 \\ \hline
\multicolumn{2}{l}{Gaussian process low activity} & $-41\pm 1$   & $-7 \pm 1 $  & $0 \pm 2$  & $-2 \pm 4 $  & $-3 \pm 1$  & & 1.75 & 1.09\\ \hline \hline
\multicolumn{2}{l}{\rm{Circular high activity }} & $-15 \pm 1 $   & $-4 \pm 1 $  & $0 \pm 2 $  & $+3 \pm 2 $  & $+3 \pm 1 $  & &  3.83 & 2.57 \\\hline
\multicolumn{2}{l}{\rm{Kepler high activity }} & $-14 \pm 1 $   & $-2 \pm 1 $  & $0 \pm 1 $  & $+4 \pm 4 $  & $+4 \pm 2 $  & & 3.35 & 2.07 \\ \hline
&\rm{ + BIS $P_{\rm{rot}}$ $3^{\rm{rd}}$ harm.}      & $-7 \pm 2 $  & $+2 \pm 2 $  & $0 \pm 2 $  & $+2 \pm 3$          &  $+4 \pm 2 $             & & 3.61  & 2.32     \\\hline
&\rm{ + $FF'$ } & $-16 \pm 4 $  & $-2 \pm 4 $  & $0 \pm 2 $  & $-1 \pm 4 $          & $-6 \pm 7 $       & & 3.75 & 2.42  \\ \hline
\multicolumn{2}{l}{Gaussian process high activity} & $-16\pm1$   & $-4 \pm 2 $  & $0 \pm 3$  & $-1 \pm 3 $  & $-3 \pm 2$  & & 2.47 & 1.57\\ \hline \hline
\end{tabular}
\end{table*}

The relative Bayesian evidences for each model with different number of Keplerians are shown in Table~\ref{tab:model_comparison_evidence}. For ease of comparison, we set the log evidence for models with two Keplerians to zero. Thus positive values are more favoured and negative ones less favoured than the 2-planet models, with significance to be interpreted according to Jeffreys' scale (Table~\ref{tab:jeffreys}). Table~\ref{tab:model_comparison_evidence} also includes the rms and median values for the RV residuals computed using the maximum a posteriori (MAP) values of the various 2-planet models. 

Note that the absolute evidences of different models cannot be meaningfully compared in a global sense as the evidence values depend, among other things, on the data fitted; hence an $n$-Keplerian model that fits only RVs cannot e.g.\ be meaningfully compared to another $n$-Keplerian model that fits RV simultaneously with a BIS or \lrhk time series. In passing we note, however, that it \emph{was} possible to compare the absolute evidence values of the parametric model `Kepler + BIS $P_{\rm{rot}}$ $3^{\rm{rd}}$ harm. + magn. cycle' with the GP model, as these fitted the same three time series; the 2-planet GP model achieved a log evidence an order of magnitude higher than the former model ($-584\pm3$ vs.\ $-4810 \pm 6$). This overwhelming difference in Bayesian evidences reflects the fact that the GP model achieved a superior fit to the observational data (RVs, \lrhk, and BIS series jointly), and moreover that high-quality fits under the GP model were localised to a comparatively small volume of prior space.

\subsubsection{Spurious third and fourth planets \label{sec:spurious}}

Table~\ref{tab:model_comparison_evidence} shows that many of the simple parametric models -- though not the GP model -- favoured the `detection' of three or even four planets. However, we have strong empirical reasons to reject models with more than two Keplerian components.

\begin{table*}
\centering
\caption{\mystar: Periods for a putative third planet found in the various runs by the different models, their corresponding semi-amplitudes and the uncertainty (per Table~\ref{tab:model_comparison_evidence}) in the model evidences. Values in bold occur in  at least 2 of the runs. }
\label{tab:model_comparison_3rdperiods}
\begin{tabular}{llccrr }
\hline 
\multicolumn{2}{l}{\rm{Model}} & $N_{\rm{runs}}$ &  {\rm{$\sigma_{\ln\mathcal{Z}}$}}& {\rm{Periods (d)}}  &  {\rm{Semi-amplitudes (\mps)}}\\\hline \hline
\multicolumn{2}{l}{\rm{Circular}}  & 5& 3&  $\mathbf{11}, 32, 63$   &  $\sim 0.6, 0.7, 0.7$\\ \hline
\multicolumn{2}{l}{\rm{Kepler}} &  3 &  4&  $ 8, 11, 19$ & $\sim  1.1, 0.7, 0.8 $\\\hline
&\rm{ + magn. cycle} & 3& 4 & $ 12, \mathbf{19}$  &  $\sim 0.6, 0.8$\\\hline
&\rm{ + linear BIS} & 3& 2 & $ 10, 12, 77$  &  $\sim 0.5, 0.5,  0.9$\\\hline
&\rm{ + BIS $P_{\rm{rot}}$ $1^{\rm{st}}$ harm.}& 3 & 7 & $11, 34, 64$ & $\sim  0.7,  0.9, 0.7$\\\hline
&\rm{ + BIS $P_{\rm{rot}}$ $2^{\rm{nd}}$ harm.} & 3& 6 & $10, 12, 34$ & $\sim  0.2, 0.2, 0.2$\\\hline
&\rm{ + BIS $P_{\rm{rot}}$ $3^{\rm{rd}}$ harm.}& 3 & 4 & $ \mathbf{12}, 26$ & $\sim 0.9, 0.9$\\\hline
&\hspace{0.7cm }\rm{ + magn. cycle}& 3 & 8 & $\mathbf{18}, 63 $ & $\sim 1.2, 1.7$\\\hline
&\rm{ + linear \lrhk} & 3& 4 & $9, 23, 29 $ & $\sim 0.7, 0.5, 0.5$ \\\hline
&\rm{ + $FF'$ }& 3 & 5 & $12, 29, 43$ & $1.1, 0.9, 0.5$\\ \hline
\multicolumn{2}{l}{\rm{Gaussian process}} & 3 & 3   & $28, \mathbf{80}$ & $\sim 0.3, 0.5$ \\\hline\hline
\multicolumn{2}{l}{\rm{Circular low activity}} & 4& 4 &  $12, 22, 29, 74$ & $\sim 0.7,  0.5,  1.1, 0.6$ \\\hline
\multicolumn{2}{l}{\rm{Kepler low activity}} & 3& 1 &  $ 10, 33, 49$ & $\sim 0.8, 0.7, 1.0$ \\\hline
&\rm{ + BIS $P_{\rm{rot}}$ $3^{\rm{rd}}$ harm.} & 3& 4 & $8, 17, 23$ & $\sim 0.2, 0.4, 0.1$\\\hline
&\rm{ + $FF'$ } & 3 &3 &  $9, 21, 65$ &  $\sim 1.0, 0.7, 0.9$\\ \hline
\multicolumn{2}{l}{Gaussian process low activity} & 3 & 4   & $9, 26, 34$ & $\sim0.5, 0.5, 0.2$ \\
\hline \hline
\multicolumn{2}{l}{\rm{Circular high activity}}& 3& 2 & $ 11, 28, 76$  & $\sim  1.2, 1.0, 1.0$\\\hline
\multicolumn{2}{l}{\rm{Kepler high activity}}& 4 &  4  & $ 12, 19$  & $\sim  1.7, 1.6$\\\hline
&\rm{ + BIS $P_{\rm{rot}}$ $3^{\rm{rd}}$ harm.} & 3& 3 & $9, 11, 34$ & $\sim  0.8, 1.3, 2.3$\\\hline
&\rm{ + $FF'$ } & 3 &4 &  $10, \mathbf{12}$ &  $\sim 1.6, 1.4$\\ \hline
\multicolumn{2}{l}{Gaussian process high activity} & 3 & 3   & $\mathbf{5}, 19$ & $\sim0.4, 2.0$ \\
\hline\hline
\end{tabular}
\end{table*}

First, we found that the periods of the third Keplerian signal (and also the fourth, where applicable), tended to change with every run, while the periods of $\sim 14$~d and $\sim 54$~d appeared as MAP values in every run, for all models with two or more Keplerians. This is demonstrated in Table~\ref{tab:model_comparison_3rdperiods}, where we list the periods of the additional Keplerian periods found by every model. The most common periods for an `extra' Keplerian were around $12$ and $19$~d -- as it turns out, these periods out are also strongly visible in the high-activity periodogram of the \lrhk\ and BIS time series (see Fig.~\ref{fig:bgls_activity}). As mentioned in the BGLS periodogram analysis, we interpret these periods as harmonics of the stellar rotation period. The fact that these periodicities manifested in Keplerian terms in all but the most complex activity models reflects the inadequacy of the simpler models at capturing all the activity variability -- neither the $12$~d nor the $19$~d Keplerian periodicities showed up under the GP model.

Second, it also turned out that these supposed planetary signals described by the Keplerians either did not show up  or their semi-amplitudes decrease to below $1$ \mps\ when considering only the low-activity subset of observations. Conversely, the semi-amplitudes of the supposed third planets occurring at $\sim 12$~d or $\sim 19$~d increased significantly when only the high activity data set was modelled compared to when using the full data set (see Table~\ref{tab:model_comparison_3rdperiods}).

Perhaps most significantly, our GP model -- which we regarded \emph{a priori} as being by far the most realistic of our activity models -- decisively favoured a 2-planet interpretation of our full set of observations. The GP modelling even favoured a 2-planet interpretation when considering only the high-activity subset of observations, and resulted in planet properties consistent with those derived from the low-activity subset or the full data set (see Table~\ref{tab:model_comparison_parameters}). This was not true for the simpler parametric models; the $FF'$ method equivocated between $2$- and $3$-planet solutions, and was not reliably able to detect the $\sim54$~d periodicity.

Our GP activity model, though relatively complex in its own right, did a comparatively good job of fitting variability in RVs unrelated to planets, when compared to most of the other activity models (see residuals in Table~\ref{tab:model_comparison_evidence}). This is despite our requirement that the GP fit activity-related variability in RVs \emph{simultaneously} with BIS and \lrhk\ time series, which ensured that the GP would be very unlikely to try to fit planetary variability. Presumably, then, under the GP modelling there was little residual activity-induced variability that could be explained using a third or fourth Keplerian term; hence these more complex models were rejected.

Regarding the parametric activity models (i.e.\ not the GP model), the addition of extra Keplerian terms beyond the 2-planet model in many cases \emph{did} enable a non-trivial fraction of extra activity-related variability to be fitted. This may have been the case, for example, because the amplitudes, phases and periods of the different terms in the harmonic models were by definition fixed, whereas the quasi-periodic GP model is flexible enough to fit both low and high activity variability, possibly with changing phases and arising from a combination of rotation periods (as might be the case due e.g.\ to differential rotation), rather than a single `master' rotation period \citep{rajpaul2017thesis}. Thus the non-GP activity models in many cases favoured solutions with three or four Keplerian terms.

The upshot, then, is that one's conclusions about the number of planetary signals present in an RV time series is extremely sensitive to whether one has modelled other, non-planetary (e.g.\ activity-induced) variability present in the RVs.

\subsubsection{Comments on computational costs}

 Given the high dimensionality of the joint stellar activity plus planetary models we considered (over $30$ parameters for some of the 4-planet plus activity models), accurate Bayesian evidence computation required a large number of posterior samples to be drawn -- in the case of $4$-planet models, several hundred million posterior samples were typically required, even with a state-of-the-art sampler such as \textsc{PolyChord}. This did not prove too burdensome for the parametric activity models; a typical run for a 4-planet model with $\sim 6$ additional stellar activity modelling parameters took $30-50$ hours on 10 cores with $2.9$~GHz clock speed each. 
 
 Evaluating a single GP likelihood, however, required among other things inversion of a $738\times738$-element covariance matrix; considering that we computed Bayesian evidences for models containing between 0 and 4 planets, and repeated these evidence calculations several times, we ended up implicitly inverting large covariance matrices many billions of times.\footnote{While a number of techniques for speeding up GP regression do exist, e.g. \citet{celerite}, we are not aware of any that are well-suited to cases where covariances need to be formulated over multiple inputs and outputs simultaneously, e.g.\ RVs jointly with activity indicators.} The upshot was that evaluating our GP models required of order fifty thousand CPU core hours on a high-performance computing platform. (Attempts to speed up computation by reducing \textsc{PolyChord}'s precision criterion led to unacceptably large scatter in computed evidences.) Accurate evaluation of the Bayesian evidences of these GP models would simply not have been feasible on a desktop or even a small computing cluster.
 
 At face value this might seem to be a serious shortcoming of the GP model. However, we note that Bayesian evidence computation is computationally challenging in general \citep{PlanetEvidence}, and it would seem that you `get what you pay for': the parametric activity models are certainly far cheaper to evaluate, though as we argued in Section~\ref{sec:spurious}, their over-simplicity inevitably leads to the detection of spurious planets. Such issues are avoided with the GP model.

\subsection{Planet characteristics}

\begin{table*}
\caption{\mystar: Comparison of the orbital parameters semi-amplitude $K$, period $P$ and eccentricity $e$ for the two planets \mystar~b and \mystar~c for each model. The parameters for the 2-planet models of the run with the highest evidence from all runs are shown here. Note that not all models favour the 2-planet case. }
\label{tab:model_comparison_parameters}
\begin{tabular}{llcccccc}
\hline
\multicolumn{2}{l}{Model} & \multicolumn{1}{c}{\rm{$K_b$ (\mps)}} &\multicolumn{1}{c}{\rm{$K_c$ (\mps)}} & \multicolumn{1}{c}{\rm{$P_b$ (d)}}&\multicolumn{1}{c}{\rm{$P_c$} (d)} & \multicolumn{1}{c}{\rm{$e_b$}} & \multicolumn{1}{c}{\rm{$e_c$}} \\ \hline \hline
\multicolumn{2}{l}{\rm{Circular}} & $3.65^{+0.26}_{-0.23}$  & $1.95^{+0.24}_{-0.22}$  & $14.1801^{+0.0011}_{-0.0014}$  &  $53.822^{+0.032}_{-0.025}$  & - & - \\ \hline

\multicolumn{2}{l}{\rm{Kepler}} & $3.62\pm 0.25$ & $2.07\pm 0.28$  & $14.17803^{+0.00098}_{-0.00067}$   &  $53.794^{+0.054}_{-0.072}$  & $0.110^{+0.043}_{-0.054}$ & $0.24\pm 0.13$ \\ \hline

&\rm{ + magn. cycle}     & $3.60\pm 0.21$ & $2.08\pm 0.23$  & $14.1781^{+0.0013}_{-0.0016}$   &   $53.801^{+0.050}_{-0.064}$  & $0.061^{+0.020}_{-0.056}$ &$0.175^{+0.066}_{-0.17}$  \\ \hline

&\rm{ + linear BIS}     & $3.54\pm 0.25$ & $2.00\pm 0.24$  & $14.1737^{+0.0012}_{-0.00097}$   &  $53.858\pm 0.059$  & $0.086^{+0.036}_{-0.066}$ & $0.140^{+0.046}_{-0.13}$ \\ \hline

&\rm{ + BIS $P_{\rm{rot}}$ $1^{\rm{st}}$ harm.}    & $3.55\pm 0.25$ & $1.96\pm 0.25$  & $14.18329 \pm 0.00070$   &  $53.745\pm 0.012$  & $0.066\pm 0.017$ & $0.212\pm 0.034$ \\ \hline

&\rm{ + BIS $P_{\rm{rot}}$ $2^{\rm{nd}}$ harm.}    & $3.61\pm 0.25$ & $2.14^{+0.25}_{-0.29}$  & $14.1780^{+0.0011}_{-0.00086}$   &  $53.778^{+0.044}_{-0.039}$  & $0.086^{+0.036}_{-0.055}$ & $0.26^{+0.13}_{-0.15}$ \\ \hline

&\rm{ + BIS $P_{\rm{rot}}$ $3^{\rm{rd}}$ harm.}    & $3.61\pm 0.24 $ & $2.09\pm 0.25$  & $14.1774^{+0.0013}_{-0.0011}$   &  $53.804^{+0.033}_{-0.052}$  & $0.089^{+0.037}_{-0.065}$ & $0.21^{+0.10}_{-0.17} $ \\ \hline

&\hspace{0.7cm} \rm{ + magn. cycle}  & $3.56\pm 0.21$ & $2.07\pm 0.20$  & $14.1789^{+0.0014}_{-0.0012}$   &   $53.803\pm 0.026$  & $0.059^{+0.024}_{-0.052}$ &$0.102^{+0.041}_{-0.099}$ \\ \hline

&\rm{ + linear \lrhk}    & $3.57\pm 0.21$ & $2.15\pm 0.23$  & $14.1782^{+0.0011}_{-0.00065}$   &  $53.777^{+0.020}_{-0.060}$  & $0.073^{+0.033}_{-0.048}$ & $0.20^{+0.10}_{-0.14}$ \\ \hline

&\rm{ + $FF'$ }  & $3.57\pm 0.26$ & $1.92^{+0.26}_{-0.34}$  & $14.1766\pm 0.0013$   &  $53.965^{+0.099}_{-0.20}$  & $0.095^{+0.046}_{-0.066}$ & $0.237^{+0.099}_{-0.20}$ \\ \hline

\multicolumn{2}{l}{Gaussian process}  & $3.67\pm 0.22$  & $2.18^{+0.22}_{-0.20}$ &  $14.1815\pm 0.0015$ & $53.753^{+0.050}_{-0.082}$ & $0.071^{+0.027}_{-0.047}$& $0.156^{+0.050}_{-0.061}$  \\ \hline \hline

\multicolumn{2}{l}{\rm{Circular low activity }} & $3.71^{+0.26}_{-0.15}$  & $2.08\pm 0.22$  & $14.1876^{+0.0049}_{-0.0089}$  &  $53.71^{+0.15}_{-0.22}$  & - & - \\ \hline

\multicolumn{2}{l}{\rm{Kepler low activity }} & $3.46^{+0.42}_{-0.36}$ & $1.65^{+0.33}_{-0.22}$  & $14.191\pm 0.016$   &  $53.81^{+0.14}_{-0.26}$  & $0.075^{+0.025}_{-0.065}$ & $0.145^{+0.044}_{-0.14}$  \\ \hline

&\rm{ + BIS $P_{\rm{rot}}$ $3^{\rm{rd}}$ harm.}    & $3.68\pm 0.25$ & $2.06\pm 0.24$  & $14.2019\pm 0.0017$   &  $53.365^{+0.024}_{-0.019}$  & $0.069^{+0.027}_{-0.066}$ & $0.160^{+0.060}_{-0.13}$ \\\hline

&\rm{ + $FF'$ }  & $3.53\pm 0.29$ & $1.88^{+0.33}_{-0.30}$  & $14.193^{+0.013}_{-0.012}$   &  $53.78^{+0.20}_{-0.28}$  & $0.058^{+0.029}_{-0.057}$ & $0.122^{+0.037}_{-0.12}$ \\ \hline
\multicolumn{2}{l}{Gaussian process low activity}  & $3.43^{+0.53}_{-0.59}$  & $1.78^{+0.40}_{-0.34}$ &  $14.1870\pm 0.0050$ & $53.89^{+0.27}_{-0.087}$ & $0.066^{+0.023}_{-0.037}$& $0.098^{+0.025}_{-0.041}$  \\ \hline \hline

\multicolumn{2}{l}{\rm{Circular high activity}} & $2.81\pm 0.40$ & $1.202^{+0.036}_{-0.058}$  & $14.18\pm 0.30$  &  $19.1^{+2.0}_{-2.5}$ & -  &  - \\\hline
\multicolumn{2}{l}{\rm{Kepler high activity}} & $2.94\pm 0.40$ & $1.41^{+1.1}_{-0.46}$  & $14.132^{+0.046}_{-0.035}$   &  $19.2^{+2.8}_{-2.2}$  & $0.187^{+0.071}_{-0.17}$ & $0.45^{+0.23}_{-0.38} $  \\\hline

&\rm{ + BIS $P_{\rm{rot}}$ $3^{\rm{rd}}$ harm.}    & $2.19^{+0.50}_{-0.43}$ & $1.20^{+0.46}_{-0.58}$ & $14.15^{-0.30}_{-0.39}$   &  $52 \pm 10$ & $0.181^{+0.060}_{-0.16}$ & $0.201^{+0.086}_{-0.19}$ \\\hline

&\rm{ + $FF'$ }  & $3.17\pm 0.33$ & $1.14^{+0.34}_{-0.52}$  & $14.155^{+0.035}_{-0.060}$   &  $53^{+14}_{-2}$  & $0.170^{+0.067}_{-0.14}$ & $0.28^{+0.33}_{-0.28}$ \\ \hline
\multicolumn{2}{l}{Gaussian process high activity}  & $2.84\pm 0.34$  & $1.04^{+0.40}_{-0.57}$ &  $14.156^{+0.065}_{-0.039}$ & $52.7^{+1.3}_{-1.1}$ & $0.150\pm 0.036$& $0.108^{+0.036}_{-0.051}$  \\ \hline \hline
\end{tabular}
\end{table*}

The posterior distributions for the inferred semi-amplitude $K$, period $P$ and eccentricity $e$ of the planets in all of our 2-planet models are summarised in Table~\ref{tab:model_comparison_parameters}.  Despite the wide array of models considered, the planetary parameters show a remarkable degree of consistency, with the credible intervals for their semi-amplitudes and eccentricities agreeing within $1\sigma$ across almost all models; the same holds true when considering only the low-activity subset of observations. Differences between orbital periods inferred under different models are never greater than a fraction of a per cent. This provide additional evidence that the detected signals are indeed planetary and robust. 
It is worth noting that the GP model favours a marginally larger MAP semi-amplitude $K_b$, and a marginally larger MAP semi-amplitude $K_c$, than the typical parametric models. However, the values favoured by the GP model are bracketed both above and below by other parametric models, and so are in no sense extreme; concern that a GP might `absorb' some of a planetary signal is clearly unfounded in this case.\footnote{After all, the framework from R15 is designed so that the GP component of the model is only able to explain variability simultaneously present in RVs and activity-sensitive time series; therefore, except in extremely pathological cases, there should be little risk of the GP wrongly fitting a planetary signal.} 
Under the parametric activity models, $K_c$ tends to \emph{decrease} in the low-activity data subset compared to the high-activity subset, suggesting that the outer Keplerian (with period broadly similar to the likely stellar rotation period) is actually being used to absorb some RV variability the too-simplistic activity models cannot account for. $K_b$ appears comparatively insensitive to stellar activity levels, which may reflect the fact that $P_b$ is about $2.5$ times shorter than the star's putative rotation period, such that little constructive or destructive interference between the Keplerian and activity signal occurs.

Putting aside the fine-grained differences between the GP and parametric activity models, Table~\ref{tab:planet_parameters} shows striking and remarkable consistency between planet parameters across almost all the models we considered, with well-constrained periods, and semi-amplitudes inconsistent with zero at $>10\sigma$ levels. On this basis we regard our modelling to represent the secure RV detection of two planets orbiting \mystar. By way of contrast, it is worth noting that even transiting planets with tightly constrained periods and orbital phases may sometimes prove difficult to characterize, with inferred semi-amplitudes diverging wildly depending on which model is used: Kepler-10~c is an excellent example \citep{{fressin2011kepler,dumusque2014kepler,weiss2016revised,rajpaul2017pinning}}. We refer, hereafter, to the planets we have detected as \mystar~b and \mystar~c with periods of $\sim 14$ and $\sim 54$~d, respectively.

\subsection{Stellar activity characteristics}

\begin{table}
\centering
\caption{\mystar: Rotation periods determined by the different models.  }
\label{tab:model_rot_period}
\begin{tabular}{llc }
\hline 
\multicolumn{2}{l}{\rm{Model}} & {Rotation period (d)}\\\hline \hline
\multicolumn{2}{l}{\rm{Kepler}} &\\\hline
&\rm{ + BIS $P_{\rm{rot}}$ $1^{\rm{st}}$ harm.}& $36.0414^{+0.0022}_{-0.0056}$\\\hline

&\rm{ + BIS $P_{\rm{rot}}$ $2^{\rm{nd}}$ harm.} & $35.2516^{+0.0045}_{+0.00087}$\\\hline

&\rm{ + BIS $P_{\rm{rot}}$ $3^{\rm{rd}}$ harm.}& $32.13212^{+0.00096}_{-0.00025}$\\\hline
&\hspace{0.7cm} \rm{ + magn. cycle} & $36.0383^{+0.0025}_{-0.00068}$\\\hline

\multicolumn{2}{l}{\rm{Gaussian process}} & $38.99\pm 0.47$ \\\hline\hline

\multicolumn{2}{l}{\rm{Kepler low activity}} &  \\\hline
&\rm{ + BIS $P_{\rm{rot}}$ $3^{\rm{rd}}$ harm.} & $36.22876^{+0.00094}_{-0.00047}$\\ \hline
\multicolumn{2}{l}{Gaussian process low activity} & $41.32^{+0.64}_{-0.92}$ \\\hline\hline

\multicolumn{2}{l}{\rm{Kepler high activity}}& \\\hline
&\rm{ + BIS $P_{\rm{rot}}$ $3^{\rm{rd}}$ harm.} & $37.93\pm 0.46$\\ \hline 
\multicolumn{2}{l}{Gaussian process high activity} & $40.0\pm 1.1$ \\\hline\hline

\end{tabular}
\end{table}

The stellar rotation period of $38.99\pm0.47$~d inferred under the GP model is broadly bracketed by the $\sim32$--$41$~d rotation periods inferred by our various parametric models (Table~\ref{tab:model_rot_period}), and agrees well with the $\sim40$~d period estimated by \citet{lovis2011}. 

The combined RV semi-amplitude ascribed to activity under the GP model (taking into account both the convective and rotational terms in the model, i.e.\ $V_c$ and $V_r$) is a little under $1$~\mps; meanwhile, the fitted RV jitter term under the GP model is of order $2$~\mps. By contrast, the `Kepler + BIS $P_{\rm{rot}}$ $3^{\rm{rd}}$ harm.' model ascribes only around $30$~\cmps\ of RV variability to activity, while absorbing nearly $6$~\mps\ of variability as white-noise jitter (see Fig.~\ref{fig:cornerplot_combo}). These differences are not surprising, given that our quasi-periodic GP function draws do not need to have constant amplitudes and phases, need not be strictly periodic, etc.: consequently, the GP can fit more activity-related variability than the more inflexible parametric model, which must instead absorb the stellar red noise via a large jitter term.

The GP model also allows us to constrain the evolution time-scale of active regions: the inferred value of $\lambda_{\rm{e}}=91\pm10$~d (Table~\ref{tab:GP_posteriors}) suggests an active region lifetime (or spot evolution time scale) of about two rotation periods, which is significantly longer than is observed for active regions on the Sun \citep{bradshaw14}. The hyper-parameter $\lambda_{\rm{p}}$ is harder to interpret physically, though the value $\lambda_{\rm{p}}\sim1$ suggests that the activity signals being modelled by the GP are only moderately more complex than a sinusoidal model, having on average something between three or four inflection points per period, compared with exactly two for a sine wave \citep{rajpaul2017thesis}. 
\begin{table}
    \centering
    \caption{MAP values and $\pm1\sigma$ credible intervals for the parameters of the activity-related parameters of the GP model.}
    \label{tab:GP_posteriors}
    \begin{tabular}{l c c}
    \hline
    Parameter & Inferred value & Units\\
    \hline
    $V_c$  & $0.48\pm 0.14              $  & \mps \\
    $V_r$  & $0.53^{+0.11}_{-0.53}      $ & \mps \\
    $L_c$  &   $0.0515^{+0.0044}_{-0.0054}$ & - \\
    $B_c$  &  $2.67^{+0.23}_{-0.32}      $ & \mps \\
    $B_r$  &  $13.8^{+2.8}_{-3.1}        $ & \mps \\
    $P_{\rm{GP}}$  &  $38.99\pm 0.47             $ & d  \\
    $\lambda_{\rm{p}}$ &  $0.996^{+0.089}_{-0.10}            $ & - \\
    $\lambda_{\rm{e}}$ & $91\pm 10             $ & d  \\
    \hline
    \end{tabular}

\end{table}

In addition to the stellar activity linked to the rotation period of the star, we are able to characterise the long-term magnetic cycle of \mystar. The two models considering this cycle `Kepler + magn. cycle' and 'Kepler + BIS $P_{\rm{rot}}$ $3^{\rm{rd}}$ harm. + magn. cycle' estimate these variations to have a period of $3682\pm 14$~d and $3681\pm 19$~d with RV semi-amplitudes of $1.61^{+0.26}_{-0.33}$~\mps\ and $ 1.77^{+0.37}_{-0.28}$~\mps, respectively. The estimate for the RV semi-amplitude of the magnetic cycle in the first model may be affected by the lack of the short-period activity, though the MAP values for both period and semi-amplitude are consistent  within their $1\sigma$ credible intervals. 

\subsection{Conclusion}

In summary, our favoured activity model is the GP model;  we might consider the  `Kepler + BIS $P_{\rm{rot}}$ $3^{\rm{rd}}$ harm.' (or arguably the $FF'$) model to be the best of the parametric activity models, even though it falls short of the GP. Both the GP and the `Kepler + BIS $P_{\rm{rot}}$ $3^{\rm{rd}}$ harm.' models favour a two-planet solution over a one-planet solution; they both strongly reject solutions containing more than two Keplerians, supporting our non-planetary interpretation for any `extra' Keplerians. In addition, they both show amongst the lowest residual RV scatters (Table~\ref{tab:model_comparison_evidence}). However, the GP model is the only one to reject under all circumstances solutions containing more than two Keplerians, and to reliably detect the second planetary signal even when considering the high-activity subset of observations. Despite its other merits, the  `Kepler high activity + BIS $P_{\rm{rot}}$ $3^{\rm{rd}}$ harm.' model has the unusual disadvantage of having locked on to what appears to be a spurious or at least unusually short stellar rotation period, unlike the GP model and some of the other parametric activity models (see Table~\ref{tab:model_rot_period}).

The MAP values and their corresponding planetary characteristics for our two favoured models are summarised in Table~\ref{tab:planet_parameters}. 
The inferred minimum masses for \mystar~b and \mystar~c are $\sim 11 \ M_\oplus$ and $\sim 10 \ M_\oplus$ 
(model dependent), and their orbital semi-major axes around the host star are $\sim 0.11$~AU and $0.26$~AU, respectively. As Neptune has a mass of $\sim 17 \ M_\oplus$, this leads us to the conclusion that both planets are likely warm Neptunes.

\begin{table*}
    \centering
    \caption{MAP values and $\pm1\sigma$ credible intervals of the planetary parameters of the two planets in the \mystar~system for the two favoured models: `Gaussian process' and `Kepler + BIS $P_{\rm{rot}}$ $3^{\rm{rd}}$ harm.'}
    \label{tab:planet_parameters}
    \begin{tabular}{l c c c c }
    \hline
    &  \multicolumn{2}{c}{\mystar~b} &  \multicolumn{2}{c}{\mystar~c} \\
         \multicolumn{1}{l}{Parameter}& BIS $P_{\rm{rot}}$ $3^{\rm{rd}}$ harm. &  Gaussian process & BIS $P_{\rm{rot}}$ $3^{\rm{rd}}$ harm. &  Gaussian process\\ \hline
        $P$ (d) & $14.1774^{+0.0013}_{-0.0011}$ & $14.1815\pm 0.0015$ &$53.804^{+0.033}_{-0.052}$   & $53.753^{+0.050}_{-0.082}$\\
        $K$ (\mps) & $3.61\pm 0.24$  & $3.67\pm 0.22$ &$2.09\pm 0.25$ &  $2.18^{+0.22}_{-0.20}$ \\
        $e$  & $0.089^{+0.037}_{-0.065}$  & $0.071^{+0.027}_{-0.047}$&  $0.21^{+0.10}_{-0.17}$  &  $0.156^{+0.050}_{-0.061}$ \\
        $M \sin(i)$ ($M_\oplus$) & $11.44\pm 0.80$ & $11.2^{+1.2}_{-0.66}       $ & $10.0\pm 1.2$ & $9.96^{+1.8}_{-0.96}$\\
        $a$  (AU) & $0.1051\pm 0.0010$ & $0.10519^{+0.00093}_{-0.0011}$ &  $0.2558\pm 0.0024$ & $0.2554^{+0.0028}_{-0.0023}$\\
    \hline 
    \end{tabular}
\end{table*}

\begin{figure*}
    \centering
    \subfigure[Kepler +  BIS $P_{\rm{rot}}$ $3^{\rm{rd}}$ harm.: 14.2~d ]{\includegraphics[width=.49\textwidth]{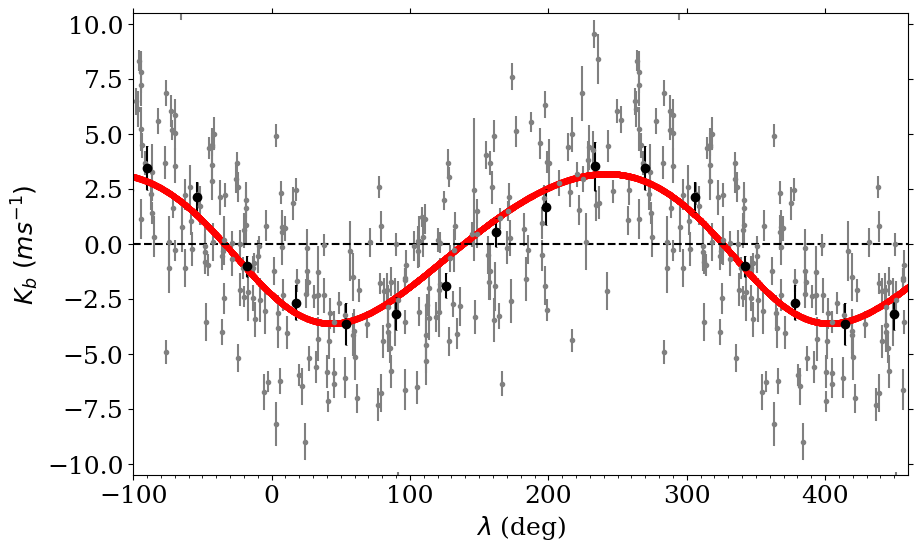}}
    \subfigure[Kepler +  BIS $P_{\rm{rot}}$ $3^{\rm{rd}}$ harm.: 53.8~d]{\includegraphics[width=.49\textwidth]{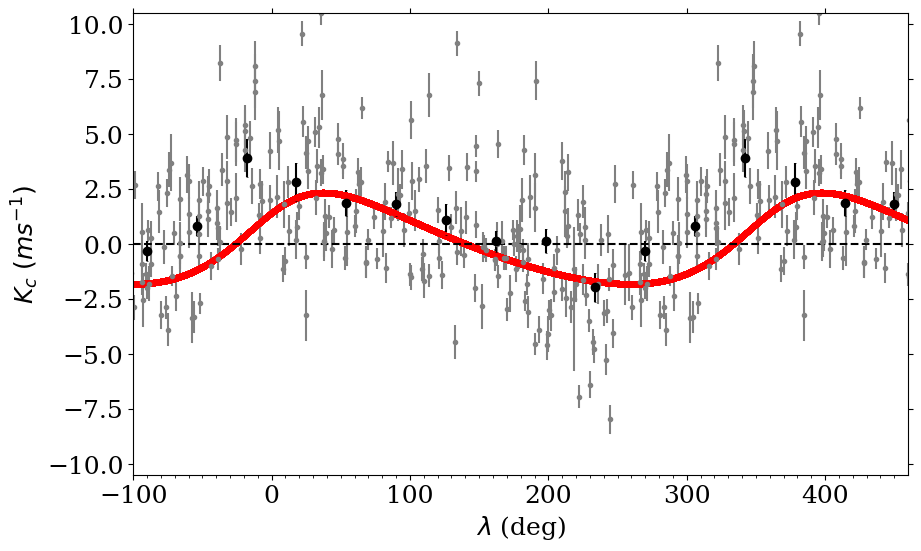}}
    \subfigure[Gaussian process: 14.2~d]{\includegraphics[width=.49\textwidth]{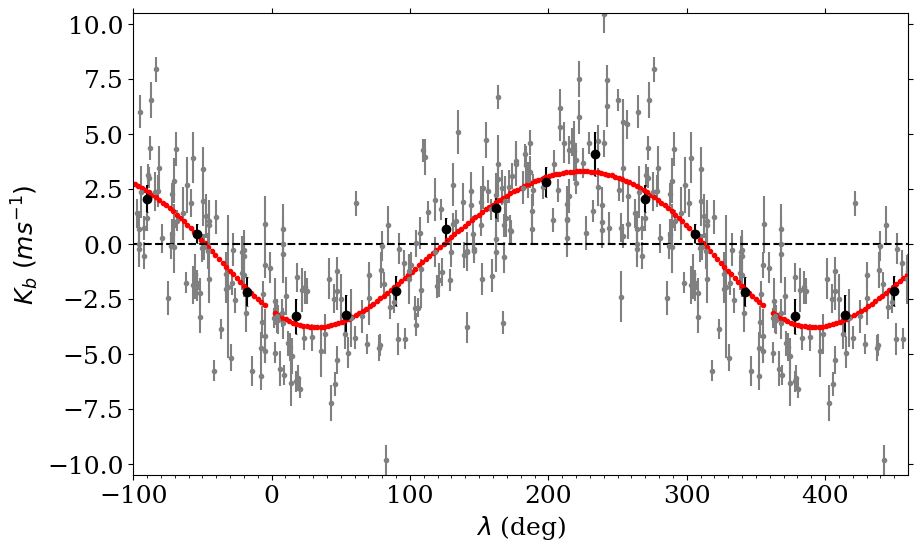}}
    \subfigure[Gaussian process: 53.8~d]{\includegraphics[width=.49\textwidth]{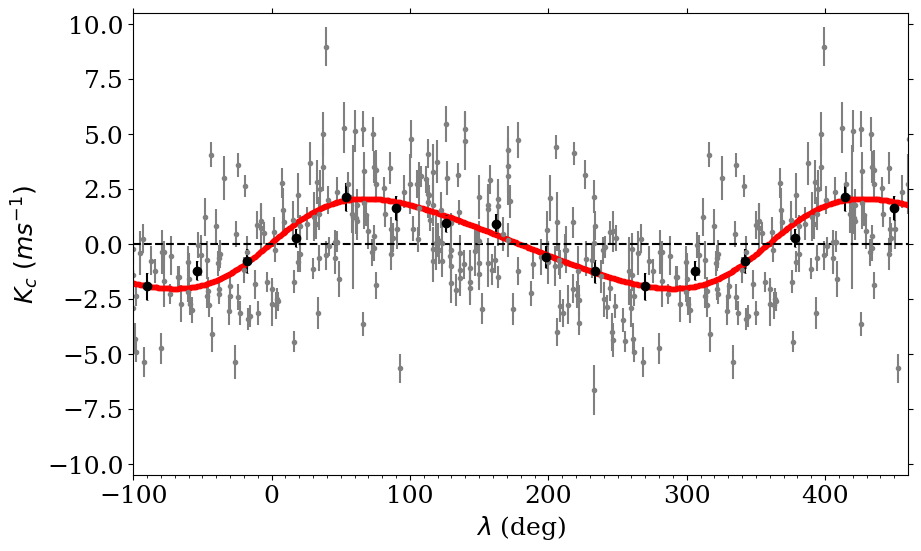}}
    \caption{Phase folded RVs of \mystar \ for the planets \mystar~b (left) and \mystar~c (right) i.e. the leftover RV when the other planetary signal, the stellar activity model and the polynomial offset is subtracted. The black dots are the averaged RV values computed over bins of 36 degrees in mean longitude and the error bars correspond to the rms of the mean, while the gray points are the observed RV values with their error. The red line represents the calculated curve of the planet with the MAP planetary parameter values of each model.}
    \label{fig:phase_folded_RV}
\end{figure*}

\begin{figure*}
    \centering
    \includegraphics[width=\textwidth]{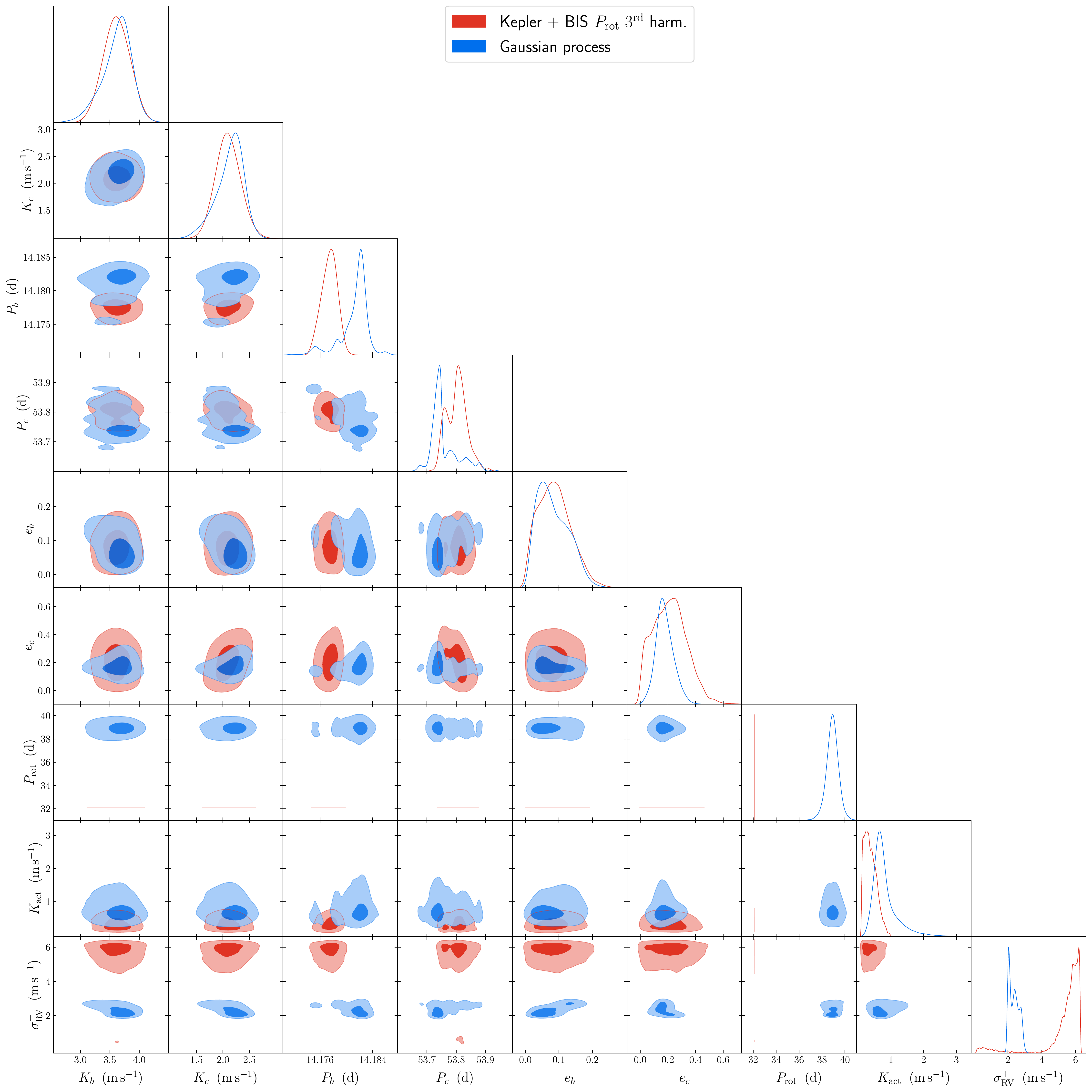}
    \caption{Corner plot of the Keplerian semi-amplitudes $K_b, K_c$, periods  $P_b, P_c$ and eccentricities $e_b, e_c$ of \mystar~b and \mystar~c, respectively, under our two favoured models: the GP model (blue) and the Kepler +  BIS $P_{\rm{rot}}$ $3^{\rm{rd}}$ harmonic model (red). $K_{\mathrm{act}}$ is the semi-amplitude of activity-dependent RV terms combined in quadrature, for the harmonic model or the GP model, and $\sigma^{+}_{\mathrm{RV}}$ is the additive noise (`jitter') parameter in either model. The dark and light filled regions correspond, respectively, to $1\sigma$ ($39.3\%$) and $2\sigma$ ($86.5\%$) joint credible regions.} 
    \label{fig:cornerplot_combo}
\end{figure*}

The phase folded RVs for both planets and both models are displayed in Fig.~\ref{fig:phase_folded_RV}, and the corner plot corresponding to the posteriors of both models is presented in Fig.~\ref{fig:cornerplot_combo}. The comparison of the posteriors of both favoured models shows that -- despite taking very different approaches to modelling stellar activity -- the planetary parameters do agree within their $1\sigma$ credible intervals. It is also interesting to note how narrow (and likely wildly over-optimistic) the posterior distribution for the stellar rotation period inferred by the harmonic model is.

\begin{figure*}
    \subfigure[Kepler +  BIS $P_{\rm{rot}}$ $3^{\rm{rd}}$ prediction]{\includegraphics[width=.49\textwidth]{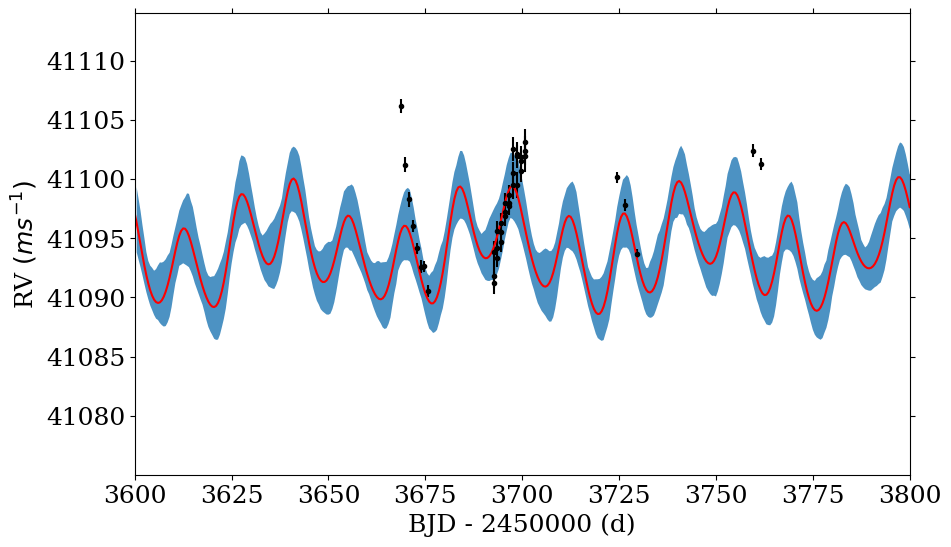}\includegraphics[width=.49\textwidth]{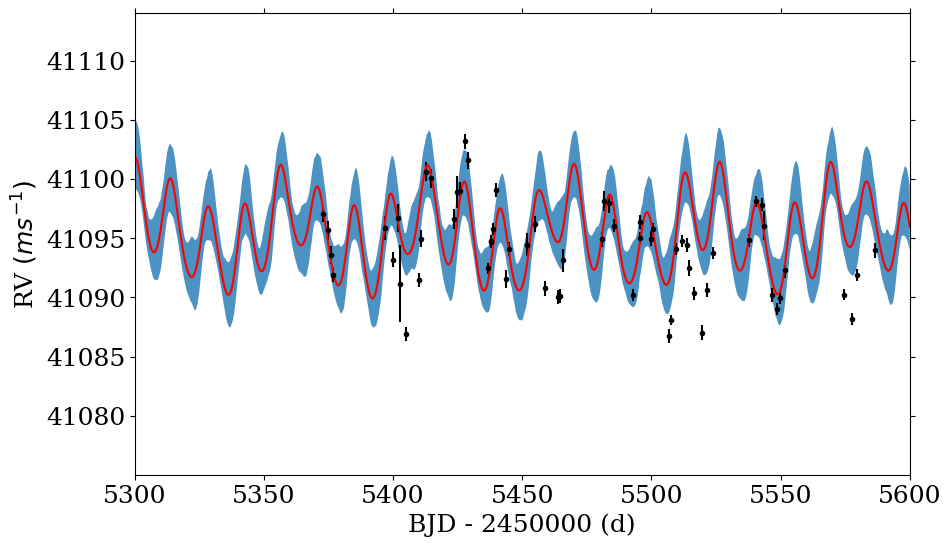}}
    \subfigure[Kepler +  BIS $P_{\rm{rot}}$ $3^{\rm{rd}}$ residuals]{\includegraphics[width=.49\textwidth]{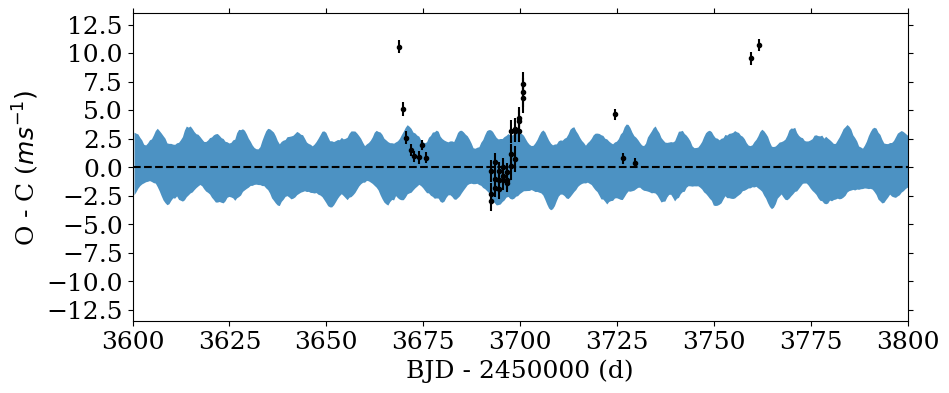}\includegraphics[width=.49\textwidth]{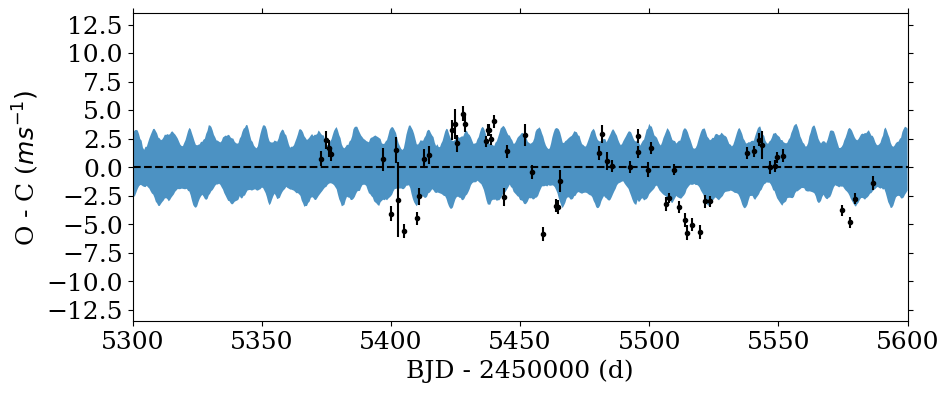}}
    
    \subfigure[Gaussian process model prediction]{\includegraphics[width=.49\textwidth]{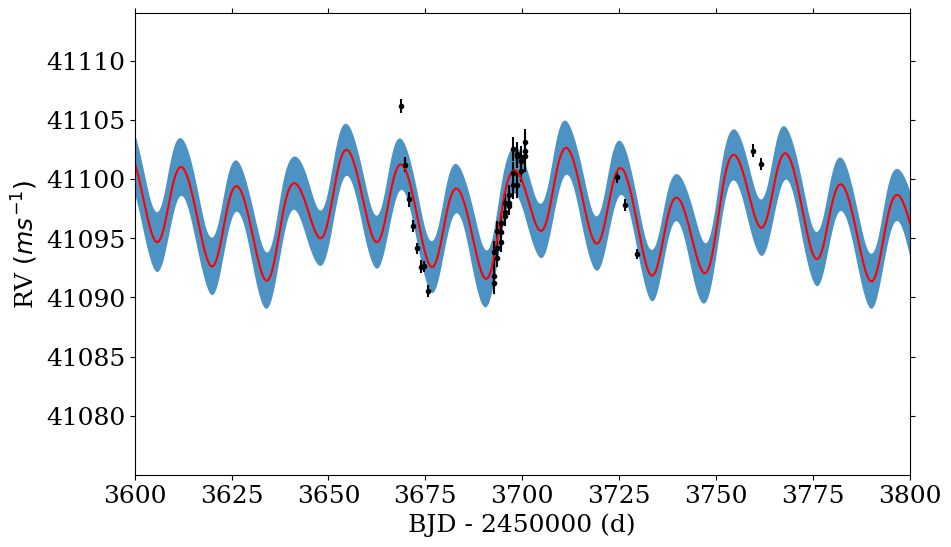}\includegraphics[width=.49\textwidth]{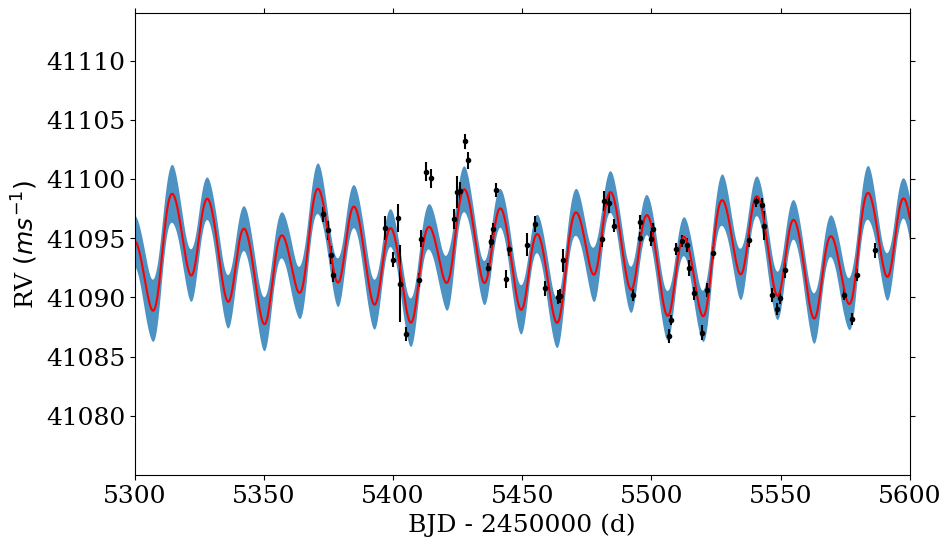}}
    \subfigure[Gaussian process model residuals]{\includegraphics[width=.49\textwidth]{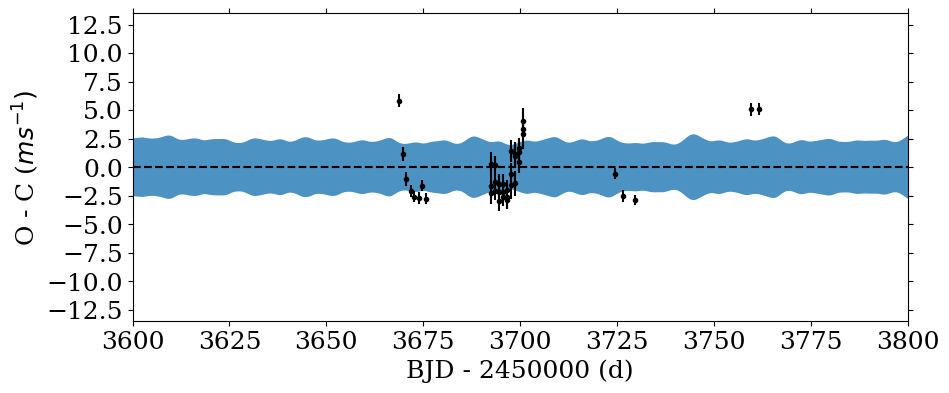}\includegraphics[width=.49\textwidth]{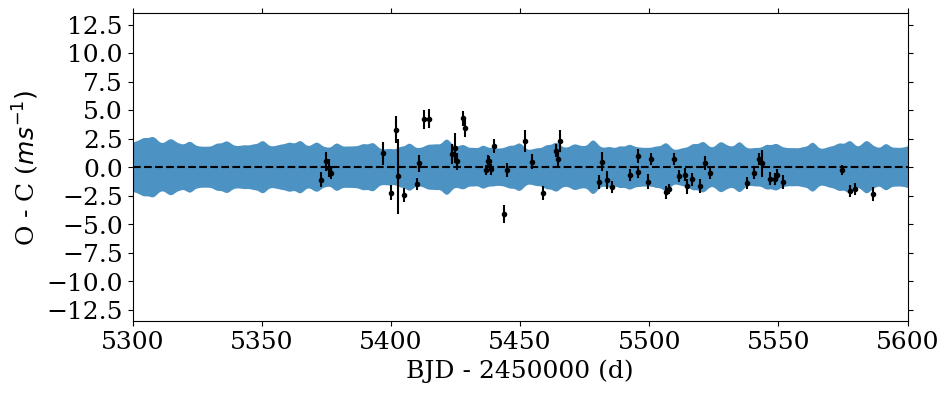}}
    \caption{Model prediction (red) with posterior predictive uncertainty (shaded blue) and their residuals for `Kepler + BIS $P_{\rm{rot}}$ $3^{\rm{rd}}$ harm.' and `Gaussian process' models; RV measurements and corresponding uncertainties appear in black. Two representative subsets of observations are shown, one from the high-activity (left) and low-activity (right) phases of the star.}
    \label{fig:model_predictions}
\end{figure*}

To demonstrate the goodness-of-fit, the RV predictions of our two favoured models in comparison to the data are shown in two parts of the data set in Fig.~\ref{fig:model_predictions}, one being in the high- and one in the low-activity subsets of observations. 

\section{Discussion}
\label{chapter:discussion}

We presented a comprehensive study of the \mystar \ RV data provided by the HARPS spectrograph. In particular, we tested and compared multiple approaches to stellar activity modelling. 

We found that Bayesian evidence comparison is, by itself, not sufficient for deciding on the number of planets present: when stellar activity is not modelled adequately, extra `planets' might be favoured to account for residual variability. Only our GP model favoured two planets (both of which we believe to be extremely secure detections) over a higher number of planets under all circumstances we considered, including modelling only a high-activity subset of observations; the parameters for the additional `planets' suggested by other models were highly variable and model-sensitive, and/or corresponded to likely activity variability (cf.\ periods in Table~\ref{tab:model_comparison_3rdperiods}, particularly for the high-activity subset). However, some of our more complex stellar activity models such as the multi-harmonic and $FF'$ stellar activity models did seem to describe the activity variability better than more simplistic parametric models, based both on the decrease in residual RV rms scatter and on the decrease in evidence for a three-planet solution vs.\ the two-planet solution. In short, model comparison using Bayesian evidence is only as good as the models considered.

Reassuringly, when modelling the full set of observations, the RV signals of \mystar~b and \mystar~c were without exception detected at a $>10\sigma$ level with \emph{every} activity model we considered, with the planets' characteristics remarkably consistent across all models. Taking all the observations into account, our two favoured models found the two planets \mystar~b and \mystar~c to orbit their host star with periods of $\sim 14.2$ and $53.8$~d at distances of $\sim 0.11$ and $0.26$~AU, with minimum masses of $11$ and $10$~$M_\oplus$, respectively.  The planetary parameters inferred from the low-activity observations alone are also mostly within agreement with those derived from the full data set. 

The influence of stellar activity on the planets' characteristics was evident, however, when considering the parameters yielded by the models applied to the high-activity data subset where the \lrhk value of the star was $>-4.90$. For example, even under a single model (`Kepler high activity + BIS $P_{\rm{rot}}$ $3^{\rm{rd}}$ harm.'), the MAP RV semi-amplitude of \mystar~b decreases from $3.71$~\mps\ to $2.19$~\mps\ when moving from the low- to high-activity subset of observations, although these two semi-amplitudes are inconsistent at only a $\sim2\sigma$ level. Using the GP -- the only model reliably to detect \mystar~c\ in the high-activity subset of observations -- the MAP RV semi-amplitude of \mystar~c decreases from $1.78$~\mps\ to $1.04$~\mps, although in this case at least the semi-amplitudes do remain consistent within $\sim1\sigma$.

To conclude, even though in our case the planetary parameter values were relatively insensitive to the choice of stellar activity model when analysing the full data set (246 measurements), caution would be essential on smaller data sets, or indeed when trying to characterise weaker planetary signals. In particular, the changes in the inferred planet parameters when considering only the high-activity subset of our data underscores the importance of stellar activity modelling even for a host star nominally classified as `inactive'.

\section*{Acknowledgements}
The authors are grateful to the anonymous reviewer for constructive feedback that helped improve this manuscript, and to Xavier Dumusque for a valuable discussion. VMR acknowledges the Royal Astronomical Society and Emmanuel College for financial support, and the Cambridge Service for Data Driven Discovery (CSD3) -- operated by the University of Cambridge Research Computing Service (\url{www.csd3.cam.ac.uk}), provided by Dell EMC and Intel using Tier-2 funding from the Engineering and Physical Sciences Research Council (capital grant EP/P020259/1), and DiRAC funding from the Science and Technology Facilities Council (\url{www.dirac.ac.uk}) -- for providing computing resources. A.M. acknowledges support from the senior Kavli Institute Fellowships.

\section*{Data Availability}
The data underlying this article will be available via VizieR at CDS.




\bibliographystyle{mnras}
\bibliography{references} 







\bsp	
\label{lastpage}
\end{document}